\documentclass[twocolumn]{aa}  

\usepackage{nicefrac}
\usepackage{thesis_defs}
\usepackage{blindtext}
\usepackage{graphicx}	
\usepackage{url}
\usepackage{epstopdf}
\usepackage{color}
\usepackage{textcomp}
\usepackage{multirow}
\usepackage{ragged2e}
\usepackage{pifont}
\usepackage{xfp,graphicx}

\usepackage[colorlinks=true,
linkcolor=blue,
citecolor=blue]{hyperref}
\usepackage{cleveref}
\usepackage{url}
\usepackage{amsmath}
\usepackage{booktabs}
\usepackage{comment}
\usepackage{afterpage}
\usepackage{mathtools}
\usepackage{todonotes}
\usepackage{makecell}
\usepackage{tabularx}
\usepackage{bold-extra}
\usepackage{xspace}
\usepackage{relsize}
\usepackage[utf8]{inputenc}
\usepackage{newunicodechar}
\usepackage[encapsulated]{CJK}
\usepackage{ucs}
\usepackage{subcaption}
\usepackage{tikz}
\usepackage{graphics}
\usepackage{upgreek}
\usepackage[utf8]{inputenc}
\usepackage[export]{adjustbox}
\usepackage{orcidlink}

\usepackage{txfonts}
\usepackage{hyperref}
\newcommand{\cmark}{\ding{51}}

\usepackage{enumitem}
\newcommand{\jr}[1]{\textcolor{green}{JR: #1}}

\begin{document}

\title{Probing circular polarization and magnetic field structure in AGN}

   \author{Joana A. Kramer \orcidlink{0009-0003-3011-0454}
          \inst{3,1*}
          \and
          Hendrik Müller \orcidlink{0000-0002-9250-0197}
          \inst{2,3}\thanks{Both first authors have contributed equally to this work.}
          \and
          Jan Röder \orcidlink{0000-0002-2426-927X}
          \inst{3,4}
          \and
          Eduardo Ros \orcidlink{0000-0001-9503-4892}
          \inst{3}
          }

   \institute{\inst{1}Theoretical Division, Los Alamos National Laboratory, Los Alamos, NM 87545, USA\\
   \inst{*}\email{jah@lanl.gov}\\
   \inst{2}National Radio Astronomy Observatory, P.O. Box O, Socorro, NM 87801, USA\\
   \inst{3}Max-Planck-Institut für Radioastronomie, Auf dem Hügel 69, D-53121 Bonn, Germany\\
   \inst{4}Instituto de Astrofísica de Andalucía, Gta. de la Astronomia, s/n, Genil, 18008 Granada, Spain
             }

   \date{\today}
   
 \abstract{The composition and magnetic field morphology of relativistic jets can be studied
 with circular polarization (CP). Recent 3D relativistic magnetohydrodynamic (RMHD) simulations coupled with radiative transfer calculations make strong
 predictions about the level (and morphology) of the jet's CP emission. These simulations show that the sign of CP and the electric vector position angle (EVPA) are both sensitive to the jet's magnetic field morphology within the radio core.}
 {We probe this theory by exploring whether the jet's radio core EVPA orientation is consistent
 with the observed sign of the core CP in deep full-track polarimetric observations. Based on a selection of sources from earlier MOJAVE observations, we aim to probe the nature of linear polarization and CP in the 
 innermost regions of jets from a small sample of nine blazars. This sample includes sources that have exhibited: (i) positive CP, (ii) negative CP, or (iii) positive \& negative CP simultaneously in the radio core region. 
 By coupling deep polarimetric observations of a carefully selected sample of blazars with state-of-the-art RMHD/radiative transfer calculations we hope to gain a deeper understanding of the physics of blazar jets.}
 {Nine blazar sources were observed using the VLBA at both 15\,GHz and 23\,GHz. Standard AIPS calibration was applied.
 Our self-calibration relies on a physically based model applied in \texttt{DoG-HiT} resulting in more accurate gains.
 To improve imaging quality, we use specialized algorithms like \texttt{DoG-HiT} that excel in handling compact emission.}
 {We observe robust, relatively high degrees of fractional circular polarization $|\bar{m_\mathrm{c}}|\simeq(0.32 \pm 0.2)$\,\% at 15\,GHz and $|\bar{m_\mathrm{c}}|\simeq 0.59 \pm 0.56$\,\% at 23\,GHz. We observe consistent polarized structure and EVPA orientation over time when comparing our analysis with archival MOJAVE data. Theoretical predictions indicate a clear favored toroidal magnetic field orientation within the extended jet emission of the reconstructed signal of the blazar 0149+218. At 23\,GHz, the jet structures of 1127-145 and 0528+134, even in superresolution, exhibit characteristics aligned with helical or poloidal magnetic nature. Changes in CP sign as frequency transitions from 15\,GHz to 23\,GHz suggest the influence of optical depth effects.}{} 

   
  \authorrunning{J.~A.~Kramer et al.}
  \titlerunning{Circular polarization in AGN Cores}
 \maketitle
%

\section{Introduction}
Supermassive black holes (SMBH) in the centers of galaxies are some of the most prominent emitters of high-energy radiation in the universe. Such objects, known as active galactic nuclei (AGN), are driven by the accretion of matter onto their central SMBH. Their emission spans across the entire electromagnetic spectrum, from radio to gamma ray energies, although only about 10\,\% are referred to as `radio-loud' AGN~\citep{Kellermann1989}.

When matter is accreted onto a BH, highly collimated plasma, called jets, outflows form along its polar axis.
Such jets are mostly visible (and studied) at radio wavelengths, identified as non-thermal synchrotron radiation, emitted by charged particles spiralling around magnetic field lines at relativistic speeds. Synchrotron radiation has the potential to be significantly linearly polarized, reaching up to 75\,\% in the presence of a uniform magnetic field~\citep{Pacho_1970,Troja}. 
Linear polarization (LP) observations
can provide valuable information about the orientation and morphology of the magnetic field structure within the synchrotron-emitting source. 
In addition, LP observations provide valuable information about the distribution of thermal electrons and the geometry of the magnetic field in the immediate vicinity of the AGN. 

Polarization in AGN was first discovered in the optical regime~\citep[e.\,g.][]{Heeschen1973} and soon also at millimeter wavelengths~\citep[e.\,g.][]{Kinman1971,Rudnick1978}. In the late 1960s, very long baseline interferometry (VLBI) 
was applied for high angular resolution studies. The first \emph{polarized} VLBI images were published in the mid-1980s~\citep{Cotton1984,Roberts1986}. 
To this day, VLBI imaging allows us to observe, resolve, and study polarized radiation emitted
from both in the innermost regions of AGN~\citep[e.\,g.,][]{EHT_M87_7,Issaoun2022,Jorstad2023} 
and their relativistic jets on the kilo-parsec (kpc) scale~\citep[e.\,g.,][]{MacDonald2017a,Hodge2018,Zobnina2023,Pushkarev2023}. 

LP is commonly expressed in terms of the electric
vector position angle (EVPA) in a VLBI image, or as the fractional polarization in some area of the jet 
with respect to the total intensity peak. Since the EVPA is predicted to be perpendicular to the local magnetic field, polarized images help us to understand the magnetic field geometry in the source. For example, extended jets up to kilo-parsecs tend to show EVPAs perpendicular to the 
direction of jet motion, indicating a poloidal or helical magnetic field. In turn, if the EVPAs are oriented parallel to the jet, the magnetic field is toroidal, 
a characteristic of shock compression~\citep{FormDisrupt2015}. A bi-modal EVPA pattern with a difference between the jet spine and sheath is indicated in theoretical models of 3D RMHD jet simulations~\citep{Kramer2021}.

Only the smallest fraction of the observed emission is circularly polarized (CP, Stokes $V$); in fact, the CP fraction only rarely even reaches 1\,\% and usually falls well below that~\citep{Wardle2021}.
Nonetheless, the use of the Very-Long-Baseline Array (VLBA) has enabled the analysis of circular polarization in extragalactic jets with exceptional precision, operating at sub-milliarcsecond resolution. Pioneering studies 
by~\cite{homan_wardle_1998} and~\cite{homan_wardle_1999} 
revealed circular polarization in the central regions of four robust AGN jets, exhibiting local fractional levels ranging from 0.3\,\% to 1\,\% of Stokes $I$ when observed with the VLBA. 

Circular polarization has subsequently been identified in other AGN jets using different instruments, as documented by~\cite{rayner_norris_sault_2000}, \cite{homan2006mojave}, and~\cite{Vitri}.
The observed CP is thought to originate either from intrinsic synchrotron processes, or from LP converted to CP by Faraday rotation~\citep[e.\,g.,][]{MacDonald2017b}. CP in AGN jets is a powerful tool for probing the particle composition and magnetic field morphology both at large scales and near the launching site. The first detection of CP in AGN jets was 
reported in the late 1990s in VLBI observations of the quasar 3C\,279~\citep{homan_wardle_1998}. 
The strongest signal detected in an AGN jet is observed in 3C\,84 by~\cite{Homan2004}.

In this work, we present the first comparison of CP maps obtained from VLBI observations with the VLBA and synthetic polarized emission maps produced with the PLUTO Code~\citep[][]{Kramer2021}. 

This paper is structured as follows:
In Sect.~\ref{sec:method} and Sect.~\ref{sec:calibration}, the methodology is outlined, including details of the 15\,GHz and 23\,GHz VLBI observations and the calibration procedures. This includes polarization calibration with particular emphasis on enhancing Stokes $V$ and rectifying compact polarized emission signals using the \texttt{DoG-HiT} imaging software.
The results are presented in Sect.~\ref{sec:Results}, where we show polarized intensity maps 
and describe the polarized structures of the nine observed blazar sources. 
A superresolution perspective on 0528+134 is also provided.
In Sect.~\ref{sec:discussion}, the results are thoroughly dissected and analyzed. This includes a view on archival MOJAVE data. Finally, Sect.~\ref{sec:conclusion} summarizes the final conclusions drawn from the resulting maps in the context of physical concepts.

\section{Methodology and observations}\label{sec:method}
\subsection{Methodology}
Over four hundred AGN jets have been observed as a part of the MOJAVE monitoring program with the VLBA from 1996 to 2016\footnote{For a  detailed listing visit the \hyperlink{https://www.cv.nrao.edu/MOJAVE/allsources.html}{MOJAVE 2\,cm Survey Data Archive.}}. From this long-term effort, the following conclusions could be drawn:
\begin{enumerate}
    \item[$\bullet$] fractional polarization in jets increases with separation from the total intensity peak and towards the jet edges of the VLBI core,
    \item[$\bullet$] 40\,\% of the VLBI cores have a preferred EVPA direction across multiple epochs, 
    \item[$\bullet$] EVPAs in jets of BL Lac objects, as well as in their radio cores, are more stable than those in quasars. Additionally, the EVPAs tend to be aligned with the initial jet direction \citep{Pushkarev2017}.
\end{enumerate}
Within the MOJAVE program, it was possible to observe CP at very faint levels within some jets (0.3$-$0.7\,\% for fractional CP, \citep{Homan2018}.
Several sources, including the blazar 3C\,279, have shown a few percent levels of CP in the radio core. A full Stokes analysis of 3C\,279 was carried out using radiative transfer to constrain the magnetic field and particle properties \citep{Homan2009}. With this approach in mind, we aim to draw our own conclusions by comparing observations and simulations of RMHD jets by
\begin{enumerate}
\item[$\bullet$] analyzing the CP dependence on the magnetic field in the VLBI core by applying the predicaments stated in \cite{Kramer2021},
\item[$\bullet$] confirming the robustness of EVPA orientation over multi-epochs by comparison to the MOJAVE archive,
\item[$\bullet$] checking whether the CP exhibits a switch from left-handed to right-handed over frequencies or time,
\item[$\bullet$] and studying the effect of various magnetic field morphologies within the extended relativistic jet emission.
\end{enumerate}

For the simplicity of our paper and our interpretation, the magnetic field structure in the sources will be assessed qualitatively by the features in polarization, in detail, by calculating the Stokes parameter and linking them to the field direction. Our goal is to analyze linear polarization $\left(\vec{P}=\vec{Q}+i\vec{U}\right)$ and circular polarization $\left(\vec{V}\right)$, and to compare it to features observed in numerical simulations~\citep{Kramer2021}\footnote{We are aware of additional affects, and we discuss opacity affects among others later in the paper. However, we choose to perform a specific comparison between our observations and models applied in~\citet{Kramer2021}.}. That is, a single sign CP is an indication for a purely poloidal magnetic field, a bimodal EVPA $\left(0.5\arctan\left(\vec{U}/\vec{Q}\right)\,\text{, pointwise division}\right)$ and two-signed CP indicates a toroidal geometry. For more details we refer to \citet{Kramer2021} and our analysis in Sec. \ref{sec:discussion}. Additionally, we evaluate the net polarization fraction, namely, net fractional linear polarization:
\begin{align}
    \bar{m}_\mathrm{l}=\frac{\sqrt{\left(\sum_i Q_i\right)^2+\left( \sum_i U_i\right)^2}}{\sum_i I_i},
\end{align}
and fractional circular polarization $\bar{m}_\mathrm{c}=-\bar{V}/\bar{I}$. Note that the net linear poalrization differs for resolved sources with a non-dominated EVPA orientation from the average linear polarization fraction:
\begin{align}
    \langle |m_l| \rangle = \frac{\sum_i \sqrt{Q_i^2+U_i^2}}{\sum_i I_i}.
\end{align}
The net polarization fraction $\bar{m}$ bears the advantage of being independent of the resolution, and allowed for more rigorous comparisons to various magnetic field configurations in simulations \citep{Kramer2021, EHT_M87_7}.

\begin{table*}[htpb!]
\small
\caption{Summary of observations}
\begin{center}
\begin{adjustbox}{max width=\textwidth}

\begin{tabular}{@{}llllllllllll@{}}
\hline
\hline \noalign{\smallskip}
IAU    & Status\tablefootmark{a} & Type\tablefootmark{b} &  \multicolumn{1}{c}{$z$\tablefootmark{c}}    &  \multicolumn{2}{c}{$\left(I^{\pm RMS}\right)$\tablefootmark{d}}  & \multicolumn{2}{c}{$\left( P \right)^{\pm RMS}$\tablefootmark{e}} & \multicolumn{2}{c}{$V$\tablefootmark{f,*}} & \multicolumn{2}{c}{Beam (bpa)\tablefootmark{g}}\\

1950.0 &    &   &   &     \multicolumn{2}{c}{(mJy/beam)}  &   \multicolumn{2}{c}{(mJy/beam)}      &   \multicolumn{2}{c}{(mJy/beam)}  &   \multicolumn{2}{c}{($\upmu$as)$\times$($\upmu$as) ($^\circ$)}    \\ \cmidrule(l){5-6}  \cmidrule(l){7-8} \cmidrule(l){9-10} \cmidrule(l){11-12}
&    &   &   &     15\,GHz   &  23\,GHz  &   15\,GHz   &  23\,GHz      &   15\,GHz   &  23\,GHz      &   \quad 15\,GHz   &  \quad 23\,GHz       \\ \noalign{\smallskip}
\hline \noalign{\smallskip}
0059+581& a & L    & 0.644 & $1685.9^{\pm 1.4}$ & $2925.24^{\pm 7.78 (7.8)}$  & $49.96^{\pm 0.53}$ & $164.61^{\pm 1.78}$ & -4.08$^{\pm 3.32}$ & -14.90$^{\pm 16.30}$ & 880 $\times$ 591 (16)& 822 $\times$ 589 (1) \\

0149+218& b &  L   & 1.320 & $362.88^{\pm 0.01}$  & $500.23^{\pm 1.33 (1.06)}$  & $6.37^{\pm 0.03}$ & $1.94^{\pm 0.03}$  & 0.51$^{\pm 0.74}$ & -1.31$^{\pm 2.70}$  & 1174 $\times$ 562 (164) &  1224 $\times$ 685 (17)\\

0241+622& a & L    & 0.045 & $980.15^{\pm 0.05}$  & $1966.26^{\pm 4.56 (3.56)}$  & $10.37^{\pm 0.66}$ & $24.01^{\pm 0.47}$ & 3.23$^{\pm 1.91}$ & -17.81$^{\pm 10.42}$  & 898 $\times$ 604 (33)& 794 $\times$ 567 (14) \\

0528+134& c & LH   & 2.070 & $524.90^{\pm 0.21}$ & $333.55^{\pm 0.02 (2.08)}$  & $9.07^{\pm 0.07}$ & $3.00^{\pm 0.04}$  & 4.91$^{\pm 1.08}$ & -2.26$^{\pm 1.89}$  & 1303 $\times$ 538 (170) & 1142 $\times$ 566 (174) \\

0748+126& b & LQ   & 0.889 & $448.06^{\pm 0.02}$ & $356.51^{\pm 0.05 (1.98)}$  & $10.17^{\pm 0.03}$ & $8.82^{\pm9.67 }$  & 2.23$^{\pm 0.90}$  & -6.55$^{\pm 1.97}$ &  1318 $\times$ 545 (176)  & 1067 $\times$ 471 (178)   \\

1127$-$145& b & LQ  & 1.184 & $690.02^{\pm 0.06}$ & $1220.17^{\pm 3.03 (1.77)}$ & $30.62^{\pm 0.14}$ & $30.79^{\pm 0.39}$ & -4.85$^{\pm 1.33}$ & 23.01$^{\pm 6.35}$ & 1365 $\times$ 548 (176)&  1196 $\times$ 370 (172)\\

1243$-$072$^\star$& a & LH  & 1.286 & $486.55^{\pm 0.31}$ & $556.61^{\pm 3.91 (2.54)}$ & $20.63^{\pm 0.36}$ & $13.11^{\pm 0.38}$ & 2.76$^{\pm 0.99}$ & 17.80$^{\pm 3.96}$  & 1376 $\times$ 562 (175) & 1198 $\times$ 353 (169) \\

1546+027$^\star$& c & LH   & 0.414 & $1068.11^{\pm 0.83}$ &  $361.11^{\pm 0.12 (7.81)}$ & $35.01^{\pm 1.78}$ & $65.24^{\pm 0.1}$  & 3.85$^{\pm 1.99}$ & 2.90$^{\pm 1.90}$  & 1290 $\times$ 562 (178)& 1470 $\times$ 455 (167) \\

2136+141& b &  LQ  & 2.427 & $552.66^{\pm 0.13}$ & $481.51^{\pm 0.21 (2.05)}$ & $11.46^{\pm 0.07}$ & $23.45^{\pm 0.09}$ & 3.65$^{\pm 1.19}$ & -38.45$^{\pm 2.89}$ & 1234 $\times$ 555 (173) & 1136 $\times$ 506 (10) \\ \noalign{\smallskip} \hline
\end{tabular}
\end{adjustbox}
\end{center}
\tablefoot{Full source list of science targets.
\tablefoottext{a}{MOJAVE status: (a) Not actively monitored. (b) Monitoring resumed. (c) Monitored at irregular cadence between monthly and yearly.} \tablefoottext{b}{Types: all sources are classified as L; LSP (low spectral peaked). Q; Quasar. H; HPQ (high polarized quasar); $|m_{\rm optical}| \geq 3\,\%$ on at least one epoch.}
\tablefoottext{c}{Redshift.}
\tablefoottext{d}{Total intensity corresponding to peak value in Figs.~\ref{fig:LP15}$-$\ref{fig:CP23} and RMS (in brackets: RMS achieved by CLEAN).} \tablefoottext{e}{Linearly polarized intensity corresponding to peak value in Figs.~\ref{fig:LP15} \& \ref{fig:LP23}.} \tablefoottext{f}{Circularly polarized intensity corresponding to peak value in Figs.~\ref{fig:CP15} \& \ref{fig:CP23} (in brackets: calibration error reported in Tab. \ref{tab:pol}).} \tablefoottext{*}{Absolute value.} \tablefoottext{g}{Beam fwhm and position angle.} \tablefoottext{$\star$}{Excluded from interpretation (see App.~\ref{app:ex}).} 
 \\ 
}
\label{tab:source_list}
\end{table*}

\subsection{Observations}
We selected a sample of AGN that have shown the characteristics of the simulations presented in~\cite{Kramer2021}, namely,
a core-focused\footnote{Defined as structure identified with a radio-core in observations or a recollimation shock in simulations, respectively} structure in CP indicative of a poloidal magnetic field
on the one hand, and a switch in the sign of CP hinting towards a toroidal magnetic field morphology on the other hand. 
The selected target sources are presented in Table~\ref{tab:source_list} along with their monitoring status, optical class, and redshift. 

The VLBA experiment BK242 was observed in dual-polarization mode, over a 24-hour period on January 6 and 7, 2022.
All ten VLBA antennas were scheduled for observation and were present for most of the allocated time. Technical difficulties at the NL, LA, KP and PT stations, as well as weather conditions at HN and BR resulted in a total of 1599\,min of downtime (11.1\,\% of the total observing time of 14380\,min, combined for all antennas). 

The data were correlated with an integration time of 1\,s at the National Radio Astronomy Observatory (NRAO) correlator located in Socorro, NM. The four intermediate frequency (IF) windows were split into 256 channels each, yielding a total bandwidth of 1024\,MHz, and containing both linear and cross-hand polarization (i.e., RR, LL, RL, LR). Observations at 15\,GHz and 23\,GHz were performed simultaneously to check the
compatibility of the results to the MOJAVE archival data. 
Furthermore, a multi-frequency dataset allows us to test the frequency dependence of structures in CP images. 

\section{Calibration and imaging\protect\footnote{All of the calibration and imaging steps detailed here were done for both observational frequencies.}}\label{sec:calibration}

\begin{figure*}
\centering
\includegraphics[width=\textwidth]{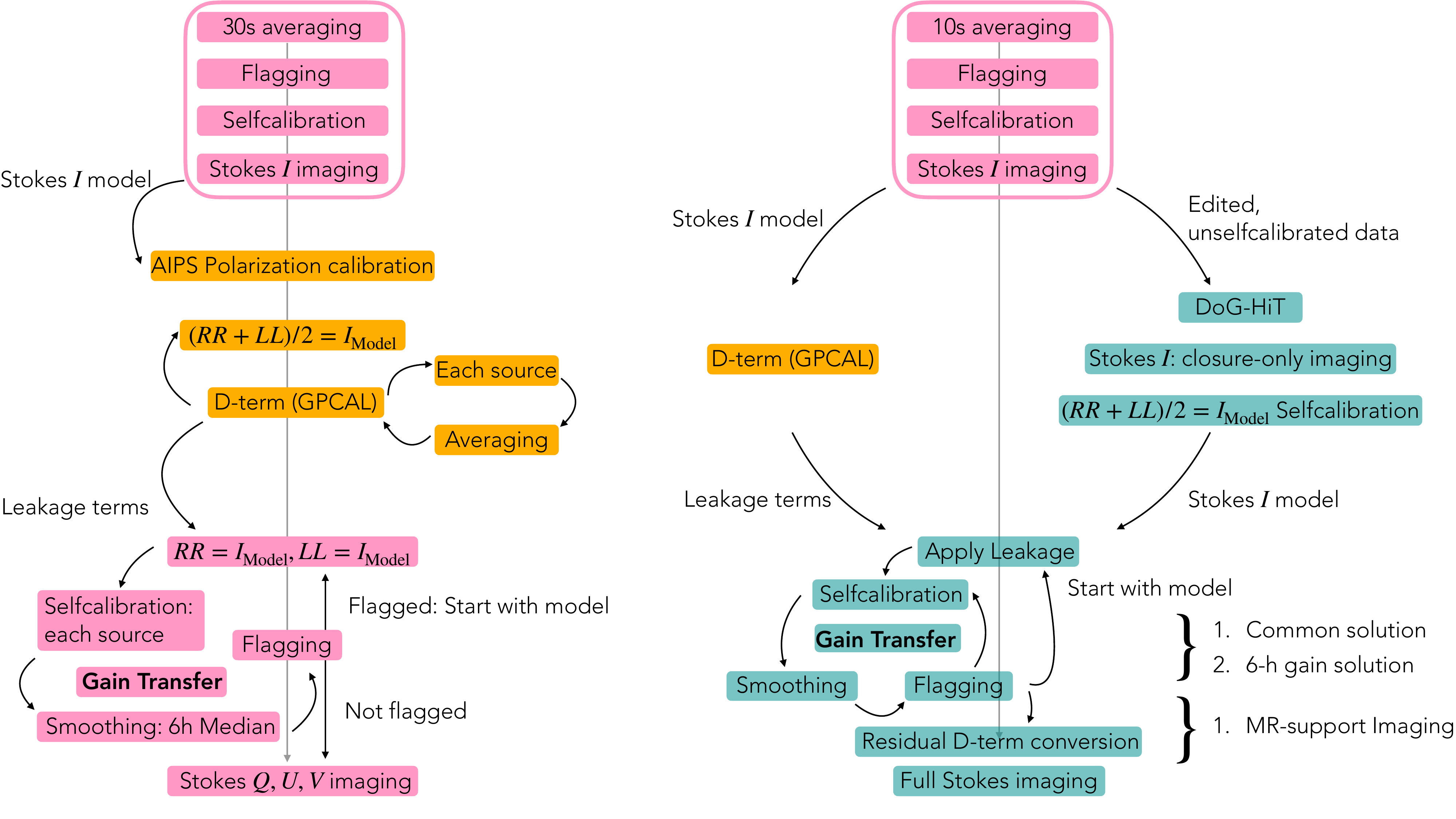}
\caption{This flowchart represents the similarities and differences between the pipeline presented in \citet[][left]{Homan2001} and our pipeline with \texttt{DoG-HiT} (right). Both pipelines illustrated here start post-calibration in AIPS. The data are processed in Difmap (pink), solved and adjusted for leakage terms in AIPS (yellow) and \texttt{DoG-HiT} in our pipeline, respectively (green), solved and applied for gain transfer, and finally imaged in full polarization.}
\label{fig:flow}
\end{figure*}

\begin{figure*}
\centering
\includegraphics[width=\textwidth]{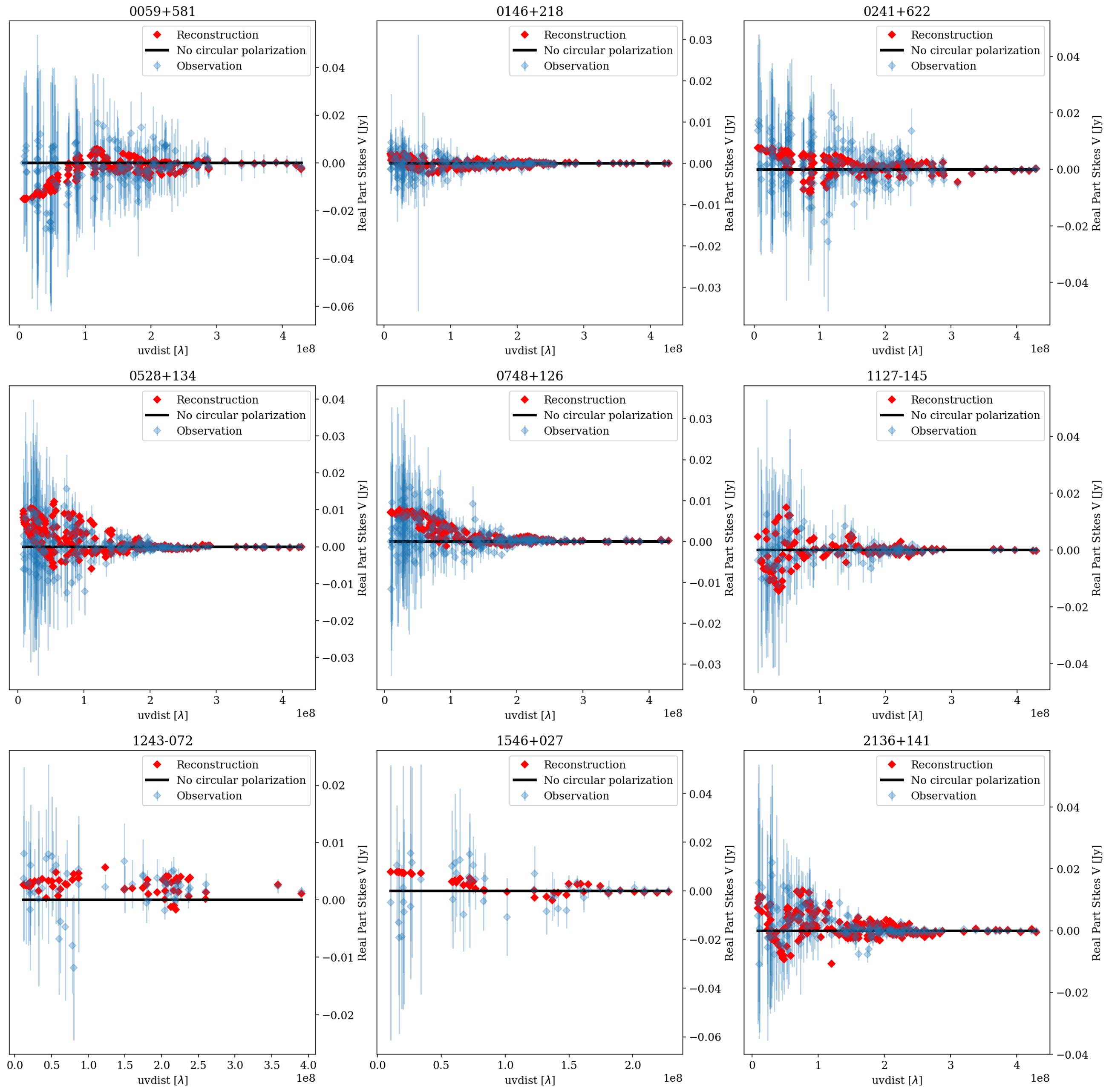}
\caption{Final calibrated and flagged Stokes $V$ amplitudes as a function of $uv$-distance (blue points) and the respective fit (red points).}\label{fig:rrll}
\end{figure*}

The circular polarization signal is challenging to recover due to the low signal-to-noise ratio, as well as the degeneracy with the total intensity image and the RL-offset calibration. 
In particular, the correct calibration of the gains and $D$-terms is crucial for robust detection of circular polarization.
The VLBI imaging process commonly alternates iterations of 
the CLEAN algorithm and a statistical self-calibration of the gains assuming a vanishing Stokes $V$ signal averaged across sources over long baselines \citep{Homan2018}. This procedure proved to result in satisfactory circular polarization maps in the past, presented e.g., in
\citet{Homan2018}. However, a CLEAN-based statistical calibration approach has some limitations: CLEAN is restricted in its resolution, uses an unphysical model of the emission, and essentially imprints these shortcomings onto the gain model.

There is an ongoing effort to develop novel imaging 
algorithms, especially inspired by the needs of the Event Horizon Telescope Collaboration \citep[see e.g.,][]{Akiyama2017a, Akiyama2017a, Chael_2016, Chael_2018_Imaging, Broderick2020, Arras2021, Arras2022, Muller2022, Tiede2022, Mueller2023a, Mueller2023c, Mus2024b}. These automated methods were found to outperform CLEAN in 
terms of dynamic range, resolution, and overall accuracy in low signal-to-noise ratio and sparse coverage settings \citep[see e.g., see the 
comparisons in][]{EHT_M87_4, Arras2021, Muller2022, Roelofs2023, Mueller2023a, Mueller2024}. In this section we 
discuss how these developments in novel approaches can be used to improve (circular) polarization 
imaging.

\subsection{Outline of calibration pipeline}
The calibration pipeline for recovering circular 
polarization from MOJAVE observations has been described and applied successfully in a series of papers  
\citep[e.g.,][]{homan_wardle_1999, Homan2004, Homan2004II, homan2006mojave, Homan2009, Homan2018}. For this manuscript, we follow the same general steps and overall ideas, but we update the pipeline at multiple stages with the modern image processing tools that have become available in recent years.

In a nutshell, the 
polarized pipeline applied by \citet{homan_wardle_1999, Homan2001, homan2006mojave} consists of several rounds of calibration and imaging. After the initial 
data 
reduction, editing and fringe-fitting of the data, total intensity imaging with CLEAN was performed and the data set were self-calibrated in the process, followed by the calibration of polarization leakage. Finally, 
the gain transfer technique was applied to calibrate the RL gain ratio, and full Stokes imaging was 
applied using CLEAN. The details of this technique are described in \citet{homan_wardle_1999, Homan2001}. The gain 
transfer technique is of particular importance for the robust reconstruction of circular 
polarization. It is based on an assumption that we will refer to as the gain transfer assumption for the remainder of the manuscript: circular polarization is 
supposedly dominated by the total intensity, and independent among sources. It is not assumed that the circular polarization vanishes in every individual source, but 
the circularly polarized signal averaged over multiple sources would amount to zero. Source-independent 
calibration is therefore desirable. The gain transfer assumption is applied by the gain transfer technique
to recover robust and smoothly varying RL gain ratios.

For the procedure that we propose here, we followed the same general steps: 
First basic calibration in AIPS, followed by total intensity imaging, leakage 
calibration, application of the gain-transfer methodology, and finally the full Stokes imaging. The outline of the
calibration and imaging pipeline is presented in Fig. \ref{fig:flow}. 
Albeit similar in many regards, our proposed calibration pipeline differs from that used by \citet{Homan2001} in several significant 
aspects.
The most important ones are as follows:
\begin{itemize}
    \item[$\bullet$] We complement the results obtained with the CLEAN algorithm by reconstructions using the \texttt{DoG-HiT} algorithm \citep{Muller2022, Mueller2023b}.
    \item[$\bullet$] The self-calibration and $D$-term calibration are aimed to be improved. In particular, we self-calibrate using a phase-less imaging routine. Moreover, the $D$-terms are obtained by comparing the leakages recovered by a standard method \citep[GPCAL, ][]{Park2021} and residual $D$-terms computed in a final step by \texttt{DoG-HiT}.
    \item[$\bullet$] Finally, the gain transfer technique is implemented by a common data fidelity functional rather than a running median.
\end{itemize}

These updates address several limitations that a CLEAN-based 
approach may face: Neither the convolved CLEAN image (which does not fit the data), nor the sample of delta components (which is not a physically reasonable description of the on-sky image) are ideal for self-calibration, as they do not describe the data and perception of the image structure simultaneously~\citep{Mueller2023a}.  
The latter might introduce systematics that tend to become frozen into the self-calibration~\citep{Pashchenko2023, JongKim2024} and hence may affect the much weaker Stokes $V$ signal. 
Self-calibration and leakage calibration compare the model visibilities with the observed visibilities and optimize the gains and $D$-terms to maximize the match between 
these quantities. This procedure would return the true gain and leakage values once the true sky brightness distribution is known. However, as explained above, neither the 
CLEAN model (representing a biased unphysical structure) nor the CLEAN image (which does not sufficiently fit the data) are well suited representations to perform the 
calibration of polarized intensities. As a consequence, performing the $D$-term calibration and gain self-calibration with the CLEAN components is expected to 
introduce small residual gain corruptions that may affect the much weaker circular polarization signal, in contrast to forward modeling techniques which directly fit an on-
sky representation of the image.

Furthermore, CLEAN is an inverse modeling approach; it is therefore impossible to separate the calibration of the RL offset or the amplitudes from the initial phase 
calibration, as well as the choices made interactively during the 
application of the algorithm. In contrast, recent forward modeling techniques relying on closure quantities 
only~\citep[e.g.,][]{Akiyama2017a, Akiyama2017b, Chael_2018_Imaging, Broderick2020, Muller2022, Mueller2023c, Albentosa2023, Mus2024b, Mueller2024b} allow to separate the phase and amplitude 
calibration, and to perform the self-calibration with an image structure which has been recovered using only the gain-independent closure quantities. Due to these limitations, we opt for the forward modeling technique DoG-HiT which may provide a more unbiased reconstruction of gains and leakages. 

The core idea behind the gain transfer technique is that when averaging across multiple sources, the 
average circular polarization should vanish. 
This concept allows to find smoothly varying RL gain ratios throughout the observation and consequently to disentangle 
the gain calibration from the circular polarization signal.
The assumption of a statistically vanishing circular polarization signal is well-motivated on short baselines. However, the long baseline structure of the Stokes 
$V$ signal, affected by, for example, local turbulence in the jet flow, may be more significant. This requires to include the Stokes $V$ structure in the calibration process, and 
consequently a more strict implementation of the gain transfer assumption. We achieve this by constructing a common data fidelity functional among multiple sources rather 
than applying a running median. We refer to Sec. \ref{sec:gain_transfer} for more details.

It has only recently become possible to overcome these issues with the current advances in imaging and polarimetric methods for 
VLBI~\citep[e.g.,][]{Akiyama2017a, Akiyama2017b, Chael_2016, Chael_2018_Imaging, Broderick2020, BroderickPesce_2020, Arras2022, Tiede2022, Muller2022, Mueller2023a, Mueller2023b, Mueller2023c, Mus2024a, Mus2024b}. In the following, we describe the \textit{individual} steps of our 
calibration methodology in more detail and verify the methodology against established methods. To this end, we present exemplary comparisons and sanity checks for the 15\,GHz 
data in the following subsections and the appendix.

\subsection{Initial calibration and imaging}
We follow the basic total intensity calibration procedure using the standard methods described in the 
\hyperlink{http://www.aips.nrao.edu/CookHTML/CookBook.html}{AIPS cookbook}. In particular, this includes corrections for instrumental delays, 
Earth orientation parameters, phase corrections for parallactic angles, amplitude corrections for digital sampling effects, fringe fitting, and solving for amplitude gain 
effects. We used Los Alamos (LA, antenna no. 5) as reference antenna when required. 

After the initial calibration with AIPS, we average the data over a 10\,s time-span, and identify and mark the data points that deviate significantly from the norm in the imaging software Difmap. We detected data issues with the Brewster and Mauna Kea antennas at 23 GHz, and consequently flagged these antennas at 23 GHz. In order for \texttt{DoG--HiT} to produce a best-fit Stokes $I$ map, 
we provide total intensity maps a-priori. For five of the nine target sources at 15\,GHz, we use data observed during the same rough time frame (January 2022) from the publicly available \hyperlink{https://www.cv.nrao.edu/MOJAVE/allsources.html}{MOJAVE data base}. 
For the remaining four target sources, and for all target sources at 23\,GHz, we create a Stokes $I$ model using Difmap iteratively improving the flagging, calibration and image model.

\texttt{DoG--HiT} enhances the Stokes $I$ image, specifically targeting compact emission characteristics.
Initially, the approach entails an unpenalized imaging round as detailed in 
\citet{Muller2022}, utilizing the CLEAN image as the foundation. Subsequently, the \texttt{DoG--HiT} 
technique is employed, resulting in the preservation of the multi-resolution support to capture key features across scales. Finally, we self-calibrate the observation to the 
reconstruction. It is worth mentioning that the \texttt{DoG--HiT} procedure produces images by closure-only imaging and in a largely unsupervised way, that is, independent of 
the initial self-calibration and human bias. 
By processing only closure quantities rather than visibilities, fewer statistically independent observables 
 are fitted, reducing the dynamic range in comparison to that achievable when fitting visibilities. However, this procedure recovers the image structure 
independently of the initial self-calibration, which has been 
identified as an important feature to ensure a bias-free circular polarization calibration further along the pipeline 
described here. To speed up the analysis and increase the accuracy, we added the $\chi^2$-metric to the amplitudes in the first round of the \texttt{DoG-HiT} procedure, and run the algorithm with a small number of iterations only to recover a first initial guess. Then this solution is used as an initial guess for the main \texttt{DoG-HiT} imaging which depends only on calibration-independent closure quantities. Moreover, \texttt{DoG-HiT} produces a physically reasonable super-resolved image that simultaneously fits the observed visibilities, hence alleviating 
possible biasing effects that may be introduced during the CLEAN self-calibration. The images presented in this work are the results of the \texttt{DoG-HiT} reconstruction 
following the additional calibration steps described below. The image structures were only determined up to a constant re-scaling factor, fixed by the initially 
CLEANed total flux and totally absorbed in the amplitude calibration. This result is based on closure only imaging. 
However, since this effect is constant on all baselines, it does not affect the relative image 
structures in any polarization channel, that is, neither the total intensity contours and the polarized signal, nor the \emph{relative} polarization fractions. 
\looseness=-1

We verified the 15\,GHz total intensity source 
structure obtained using \texttt{DoG-HiT} (Fig. \ref{fig:15doghit}) with CLEAN images of all sources (Fig. \ref{fig:15clean}), both blurred to the same resolution.
Apart from the 
absence of CLEAN artifacts, \texttt{DoG-HiT} reconstructs individual features within the jets more 
clearly than they appear in CLEAN. All nine total intensity maps of the 
observed sources are in overall good agreement between imaging methods. For more details, we refer to Appendix \ref{app:comparison}.

\subsection{Leakage calibration}

After the cross-hand delay calibration using the task RLDLY in AIPS, the $D$-terms were finally 
calibrated during the full-Stokes imaging in \texttt{DoG-HiT} 
\citep{Muller2022, Mueller2023a, Mueller2023b}. 
We solved for leakages in a two-step procedure: First, we obtained initial $D$-terms from the prior CLEAN images using GPCAL, and then solved for residual $D$-terms with \texttt{DoG-HiT} in an iterative manner.

The initial $D$-terms were estimated by the pipeline \texttt{GPCAL} \citep{Park2021} which applies AIPS tasks \citep{Greisen_2003}. We applied the pipeline on three target sources and found the $D$-term calibration IF by IF with typical amplitudes of up to $\pm 2\%$. 

In Fig. \ref{fig:dterms} we present the final leakages recovered by \texttt{DoG-HiT} compared to the ones recovered by GPCAL. The $D$-terms, both right-handed and left-handed, are rather small
and match with reasonable accuracy. However, we also see slight differences between the methods, probably (but not necessarily) due to the possible bias within the CLEAN self-calibration procedure due to the use of a point-source model.

\subsection{Gain transfer}\label{sec:gain_transfer}
The optimal Stokes $I$ model for each source, along with the modified/unself-calibrated 
data, served as input for \texttt{DoG--HiT}. The pipeline involves resolving and fine-tuning compact polarization through the implementation of the \texttt{DoG--HiT} 
algorithm. To initiate the process, a-priori $D$-term solutions obtained from GPCAL are 
applied. Instances of nonphysical high circular polarization are appropriately flagged 
for attention. Subsequently, both amplitudes and phases are calibrated within a 6-hour time frame. 

During the calibration, we assumed the difference between right and left circular polarization $\langle \mathrm{RR}-\mathrm{LL}\rangle$ to be close to zero, indicating a small CP, vanishing statistically when averaging among various sources (but not necessarily on every source indidually) -- consistent with the strategy applied in, for example, \citep{homan2006mojave, Homan2018}. 
\citet{homan2006mojave} performed the calibration of the RL gain 
ratio on every source individually, also assuming a vanishing circular 
polarization signal. 
It is, however, impossible to separate the circular polarization signal from the self-
calibration. Therefore, the gain solutions were smoothed across multiple sources by a 
running median over six hours. The calibration procedure is as follows: 
\textit{i)} iterative calibration, \textit{ii)} flagging of bad data points (strong 
polarization), and \textit{iii)} smoothing the calibration tables with a running median. 
This strategy implements the gain transfer assumption: the circular 
polarization signal of a single source is degenerate with the calibration procedure, and we 
can make the reasonable assumption that \textit{averaged across multiple sources the 
circular polarization vanishes}.

For this manuscript, we attempt to implement this 
assumption more strictly. Due to the recent development of 
methods that modularly realize VLBI imaging and calibration by, for example, convex optimization (\texttt{ehtim}, \texttt{MrBeam}, 
\texttt{resolve}) we can easily define a common data 
fidelity functional. We aimed to find a single solution for 
the RL-offset by fitting \textit{one} (smoothly varying) solution to all data 
sets, that is, to a combination of the single data fidelity 
metrics of the single sources. The gain solution was computed with a correlation 
time length of 6 hours. To this end, we solved the minimization problem:
\begin{align} \nonumber
    g^r_i(t), g^r_j(t) &\in \mathrm{argmin}_{g_i, g_j} \\ \nonumber
    &\norm{\int_{t-\Delta t}^{t+\Delta t} V^{RR}_{ij}(\tau)-g_i^* g_j \mathrm{FFT}(\vec{I})([u_{ij},v_{ij}](\tau)) d\tau} \\ \nonumber
    g^l_i(t), g^l_j(t) &\in \mathrm{argmin}_{g_i, g_j} \\
    &\norm{\int_{t-\Delta t}^{t+\Delta t} V^{LL}_{ij}(\tau)-g_i^* g_j \mathrm{FFT}(\vec{I})([u_{ij},v_{ij}](\tau)) d\tau}
\end{align}
Here $V^{RR}$ and $V^{LL}$ denote the righthanded and lefthanded parallel hand visibilities, $g^{r/l}$ the gains, $i,j$ are indices counting the antennas, and $\Delta t$ is the time-scale of the time average, i.e., 6 hours. In contrast to self-calibration performed on every scan, we therefore aim to recover gain solutions that are consistent with the data over a long interval of times, and sources. In consequence, the gain curves are smoothed, similar to the effect achieved by a running median.

We show an exemplary set of amplitudes before application of the gain transfer technique in Fig. \ref{fig: amplitude_jump} and the gain curves for all four IFs for three randomly selected antennas in Fig. \ref{fig:rlcurve}. The gain curves look reasonably smooth, as expected. However, we detect some larger phase swings at the edges of the observing window. This may be explained by the fact that the smoothed minimization is more challenging to compute at the edges of the observing window with less neighboring estimations that can be used to construct a running median/smoothed gain curve. These times have thus been the target of more thorough flagging, as outlined below.

We likewise computed the corrections iteratively. We fitted a gain solution to the data sets implying temporal smoothness, investigated the goodness of fit, and flagged bad data points. Then, we refined the gain solution and proceeded until convergence was achieved. We would like to highlight however, that the gain curve after each flagging is always applied to the leakage-calibrated original, non-R/L calibrated data set now with additional flags. Hence, the gain curve is only applied once for the final data set used for full Stokes imaging. The need for flags at this step in the analysis stems from multiple motivations. First, we performed closure-only imaging in the previous step with \texttt{DoG-HiT}. The typical iterative mapping, flagging and self-calibration done interactively with CLEAN was therefore not applied which leaves a bigger need to do this flagging at latter steps of the analysis. Second, as outlined above, at the edges of the observing window, we need to apply extra caution to the gain solutions. This flagging has been performed manually. In Fig. \ref{fig: amplitude_jump}, we present a typical baseline that has been flagged for early times. We discuss the heuristics of our flagging procedure in more detail in Appendix \ref{app:comparison}.

Finally, we find a single, temporally smoothed gain curve that 
reflects the best compromise in fitting the individual data 
fidelities with this procedure. We present the gain curves per IF for three randomly selected antennas in Fig. \ref{fig:rlcurve}.
Our final circular polarization images in both 
frequencies are presented in Figs. \ref{fig:CP15}, \ref{fig:CP23}.
\looseness=-1

For comparison (and to validate the pipeline), 
we also calibrated all sources using the strategy described 
in detail in \citet{homan2006mojave}. The resulting 15\,GHz circular polarization maps are shown in Fig. \ref{fig:CP15homan}. The results are remarkably similar to those obtained by \texttt{DoG-HiT}, with notable exceptions. 1243$-$027 has a bi-modal core structure in circular polarization, which is rarely reflected with the second calibration technique.
This shows that the exact structure of these bi-modal 
signatures is uncertain, and should not be over-interpreted.
Further, some significant differences can be seen for 1127$+$145 and 1546+027 where the circular polarization signal appears less structured and noisy when using \texttt{DoG-HiT}.

In our sample at 15\,GHz, only one out of nine sources has a negative sign. This may point towards a potential issue during the application of the gain transfer technique, i.e., its basic assumption may be violated. As an additional validation test, we inspected this behavior and have observed that incorporating longer baselines increased the average circular polarization deviating from zero. This occurs because self-calibration incorporates all baselines, and some sources (e.g., 0149+218, 0528+134, 0748+126, 1127-145) are resolved. Our imaging process, being more sensitive to small-scale structures, differs from classical CLEAN, and longer baselines contribute significantly to the gain calibration as well. Thus, the assumption that CP averages to zero across sources applies not only to short but also to long baselines.

To validate this, we flagged the longest baselines (larger than $0.2\,G\lambda$) during gain transfer and re-imaged at 15 GHz. This produced a more uniform CP distribution (four sources negative, five positive). These results, along with smoothed gain curves for Fort Davis, Kitt Peak, and St. Croix, are detailed in Fig. \ref{fig:rlcurve}. While most gain curves were consistent, flagged data sets showed smoother gains, particularly reducing the edge-related issues in potentially problematic time intervals mentioned above.

We show our final CP maps when using these gains in Fig. \ref{fig:tapered_cal}. For most sources, the relative CP structures remained consistent to the level of showing the same structural features (e.g., single-peaked or double peaked, and trend of CP along and transverse to the jet), except for 1243 and 1546, which varied significantly across calibration methods. This suggests their CP signals may be influenced mainly by gain calibration. Notably, 1243 and 1546 had the sparsest uv-coverage, making them the weakest constrained sources.

\subsection{Full Stokes imaging}
Since gain-- and $D$-term calibration, and particularly the corresponding flagging, may have affected the Stokes $I$
observables, we refine the total intensity imaging. We refine the total intensity images with a small-stepsize gradient descent algorithm fitting the flagged and calibrated data set, starting from the earlier computed Stokes $I$solution.
Following this, the linear polarization imaging is subjected to the \texttt{DoG--HiT} 
procedure as described in \citet{Mueller2023b}.

The imaging process is further extended to address circular polarization using 
\texttt{DoG--HiT} in the same way, that is, by fitting the Stokes $V$ 
visibilities, but only 
varying the parameters in the multiresolution support \citep[following the philosophy in][]{Mueller2023b}. That means, we are only using the multiscale basis functions that were found to be of statistically significant to represent the total intensity to model the linear and circular polarization.

A five-round iterative refinement cycle is established for linear and circular polarization, and residual $D$-term calibration. This comprehensive approach aims to progressively enhance the precision and quality of the (linear) polarization results.
\looseness=-1

Residual RL offsets may still be present in the data even after these imaging and calibration steps, especially on 
long baselines where the gain transfer assumption is potentially 
inappropriate. This can result in pointy, noisy reconstructed circular 
polarizations, especially when recovering images at super-resolution. We refined the 
imaging of Stokes $V$ with the de-noised, compact Stokes $V$ 
reconstructions. To this end, we implemented manual procedures in \texttt{DoG--HiT} that are described in the following paragraphs. 

Based on the 1\%-contour of the Stokes $I$ image, a mask delineating a 
compact and luminous region was created. 
This mask was then applied to the $Q$, $U$, and $V$ 
polarization components. Moreover, the Stokes $V$ structures were blurred using a Gaussian filter. 


Following these adjustments, the final images were generated by minimizing the $\chi^2$-metric to the fully calibrated Stokes $V$ 
visibilities with a gradient descent approach stopped by the discrepancy principle starting from the blurred, and masked circular polarization images created in the last step. To this end, we allowed only pixels within the $1\%$ mask to vary.

In Fig. \ref{fig:rrll} we show the final calibrated circular polarization 
amplitudes and image fits. The observations are fitted adequately with smooth (regularized) solutions, indicating a successful fit given the various constraints imposed on the polarimetric deconvolution.

The final images show a range of anti-symmetric peaks in CP which may be introduced to the image by phase calibration errors. \citet{homan2006mojave} detected similar structures, and proposed to apply a last phase self-calibration step at the end to remove them from the images, assuming $V^{RR}, V^{LL}={\rm FFT}(I)$. To test whether this is the case, we applied this strategy as well, and show the reconstructions after a final self-calibration step in Fig. \ref{fig:CP15selfcal} and Fig. \ref{fig:CP23selfcal} as well.

Further, we would like to note that a CLEAN-based pipeline may be also prone to over-align phases in general. This is because the phase calibration cannot be separated from the amplitude calibration, and the initial phase calibration is necessarily performed during the Stokes $I$ imaging on a starting model, or a model with only a few components. That may introduce phase solutions that are too much aligned on a rather simplistic symmetric structure early on the analysis, something that may be challenging to correct afterwards during the gain-transfer. Moreover, we selected sources that were expected to show rich CP.

\subsection{Uncertainty Estimation}
In Table \ref{tab:source_list} we present the achieved RMS noise for the reconstructions with \texttt{DoG-HiT} for a comparison with literature results. We note that the \texttt{DoG-HiT} procedure does not minimize a residual in the classical sense (but fits data properties by forward modelling), and works with the closure quantities rather than the visibilities (which do not form a dirty map/dirty beam deconvolution pair due to the non-linearity of the forward operator). To define a quantity akin to the residual RMS commonly used in VLBI, we here report the ${\rm rms(FFT^{-1}}[V-{\rm FFT}(\vec{I})])$ , where $I$ is the recovered image, and $V$ are the observed visibilities. However, we would like to note that potential over-alignment of the phases to the closure-only fit, the extent of flagging, and the explicit involvement of gridding uncertainties (that are still present in the CLEAN residual, and are typically drastically reduced by multiple major loop iterations) affect the RMS. Moreover, overfitting of the noisy structure would show up as a reduced noise-level, with the noise-structure imprinted in the forward model. We have evaluated our images visually by eye to assess that latter did not appear, with the notable exception of 0149+218 at 23\,GHz which shows an extended structure around the core-jet structure that is most likely attributed to imaging or calibration artifacts rather than true image structure. In this case, we rather applied the peak of non-true image structure rather than the RMS. All these drawbacks should be taken into account when comparing to noise levels reported by CLEAN. At 15\,GHz we typically achieve an RMS level of $\sim 0.1\,\mathrm{mJy/beam}$ which lies well in the ballpark that may be expected by VLBA observations at this frequency, with 0059+581 as notable exception. For 23\,GHz, we typically score RMS errors of a few mJy/beam, leading to dynamic ranges of $\sim 500-2000$. We note more variety across the sources in the RMS at 23\,GHz, and therefore present the RMS achieved by the CLEAN data reduction done in the first step of the data analysis as well for comparison. The RMS error with \texttt{DoG-HiT} may be underestimated, especially for 0528+134.

\begin{figure*}
    \centering
    \includegraphics[trim={0cm 0cm 0cm 4cm},clip,width=\textwidth]{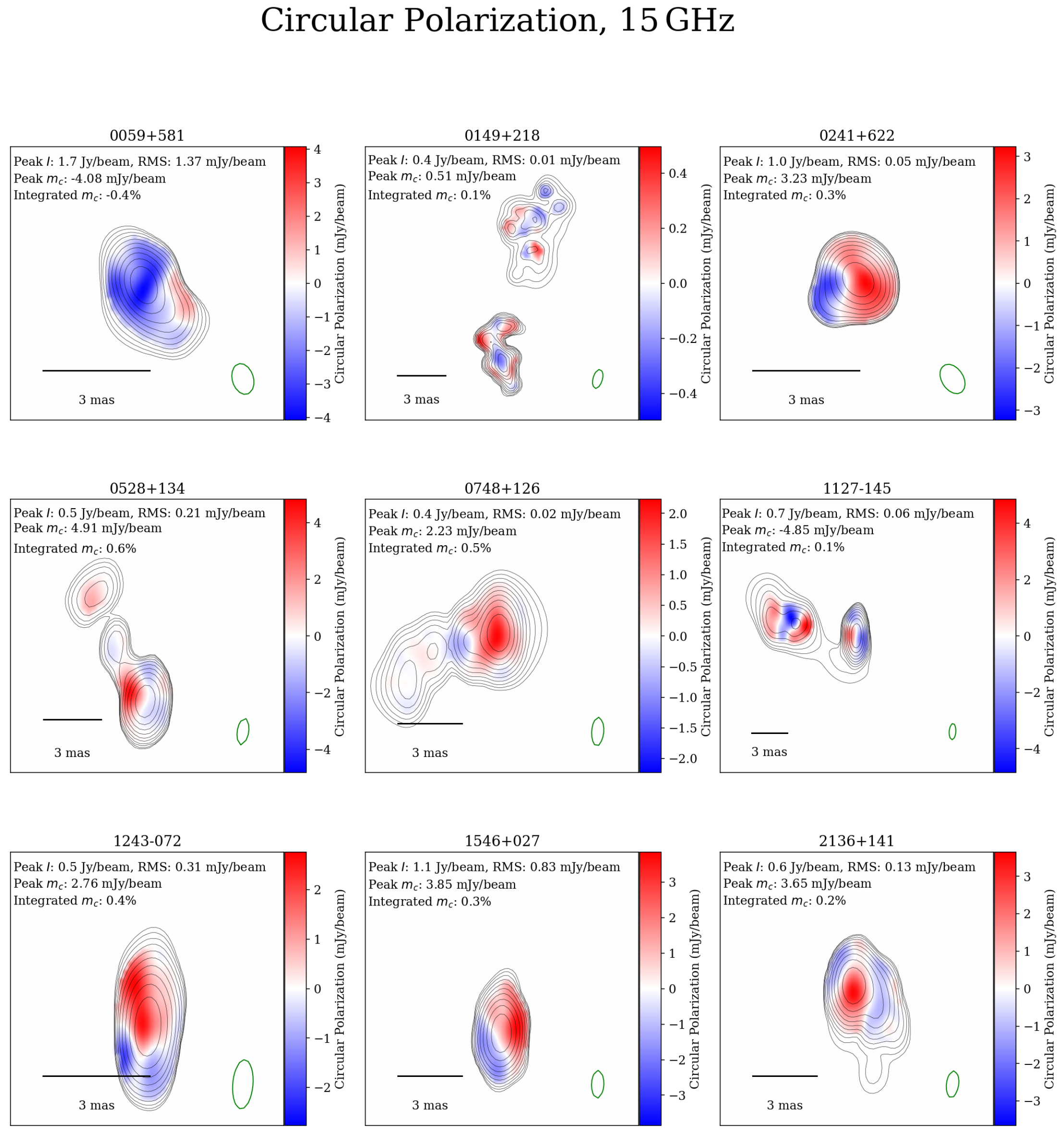}
    \caption{Circular polarization imaging results from 6-7 Jan. 2022 for nine target sources: 0059+581, 0149+218, 0241+622, 0528+134, 0748+126, 1127$-$145, 1243$-$072, 1546+027, and 2136+141. The figure shows maps for each source in circular polarization at 15\,GHz (colorcoding: blue/red negative/positive). The total intensity is indicated in white contours on each map at levels of $[0.1\%,0.2\%,0.4\%, ..., 51.2\%]$ of the peak brightness. An individual convolution beam is shown in the lower right corner of each map. The field of view and source size is shown by the scale with $3\,$mas in the lower left.}
    \label{fig:CP15}
\end{figure*}

\begin{figure*}
    \centering
    \includegraphics[trim={0cm 0cm 0cm 4cm},clip,width=\textwidth]{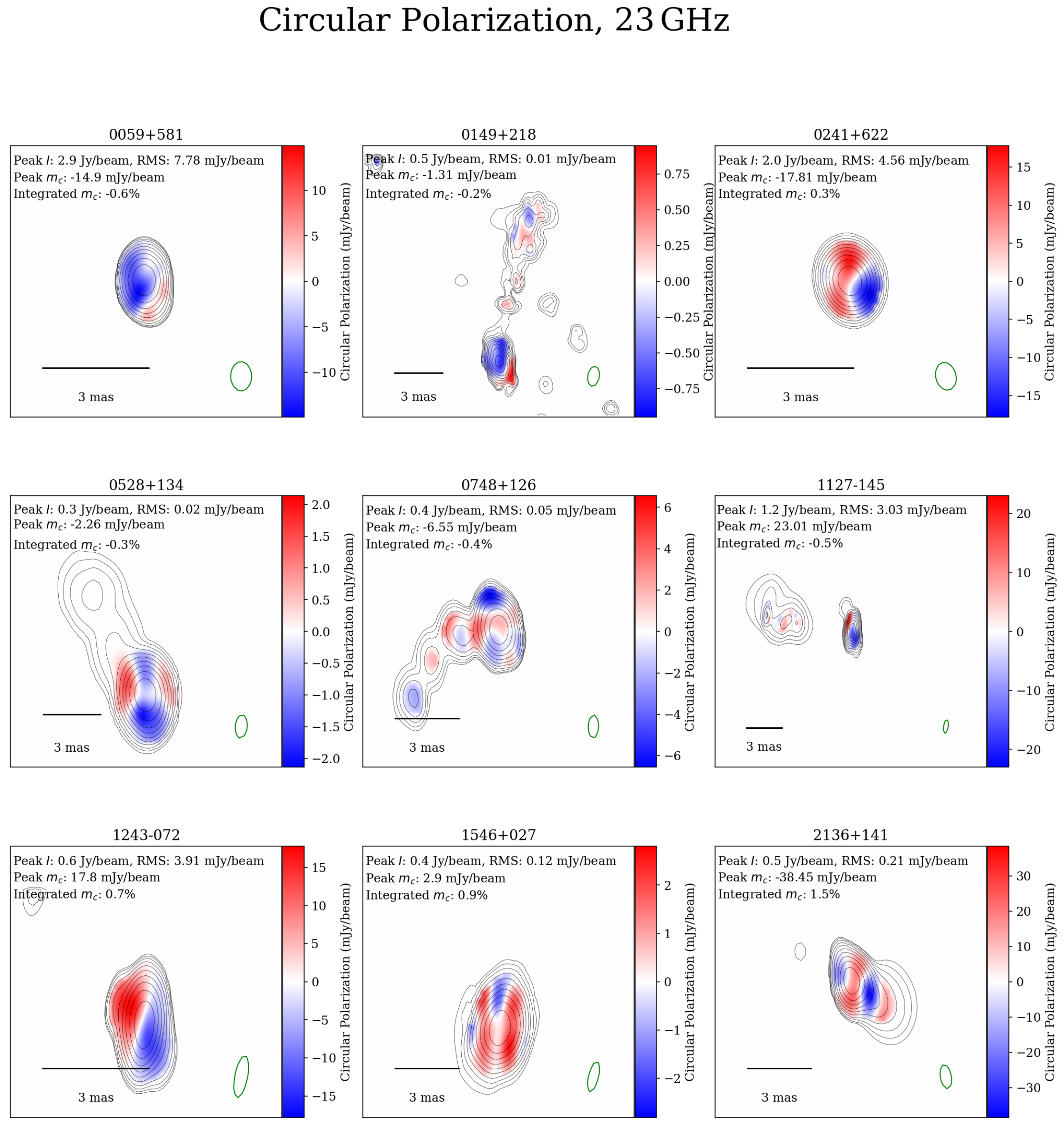}
    \caption{The same as Fig.~\ref{fig:CP15} but at 23\,GHz observational frequency.}
    \label{fig:CP23}
\end{figure*}

The CP reconstruction is strongly affected by the gain calibration. A first idea of the relative uncertainty of the recovered features may be available by comparing the recovered features with different calibration techniques, i.e., by comparing Fig. \ref{fig:CP15}, \ref{fig:CP15homan}, Fig. \ref{fig:tapered_cal} and Fig. \ref{fig:CP15selfcal} at 15 GHz, and Fig. \ref{fig:CP23} and Fig. \ref{fig:CP23selfcal} at 23 GHz.

For an analytic approximation to the errors in the CP reconstructions, we apply the scheme proposed by \citet{Homan2001}. The residual RMS of the CP reconstruction is typically underestimating the true uncertainty due to the uncertainty in the gains. \citet{Homan2001} therefore suggested to estimate the uncertainty from three components: the uncertainty in the determination of the smoothed gains quantified by the scan-to-scan variations of the gain solutions across all sources, the uncertainty in true circular polarization of the calibrator (estimated by the RMS of the apparent CP), and finally uncertainties introduced by uncorrected rapid variations in the gains (estimated for every source by variations of the gain solutions in time). For more details, we refer to \citet{Homan2001}. Note that our calibration technique differs in some details from the calibration technique this uncertainty estimation was tailored to. In particular, we originally do not compute gain solutions at every scan that we smooth in post-processing, and that could be used to estimate the scan-to-scan variations for the uncertainty estimate. We rather try to fit a smooth curve directly, compare our discussion in Sec. \ref{sec:gain_transfer}. However, the two approaches are philosophically similar. Therefore, we recomputed the gain solutions and running median calibration as proposed in \citet{Homan2001} at both frequencies and adapt the related error estimates also for our results.

\begin{table}[]
\small
\caption{Fractional polarization}
\begin{center}
\begin{adjustbox}{max width=0.5\textwidth}

\begin{tabular}{@{}lcccccc@{}}
\hline
\hline
\noalign{\smallskip}
IAU &  \multicolumn{2}{c}{$\bar{m}_\mathrm{l}$\tablefootmark{a}}   & \multicolumn{2}{c}{$\bar{m}_\mathrm{c}$\tablefootmark{b}} & \multicolumn{2}{c}{$\sigma_\mathrm{gain}$\tablefootmark{c}}\\

& \multicolumn{2}{c}{(\%)}  & \multicolumn{2}{c}{(\%)} & \multicolumn{2}{c}{(\%)}\\
\cmidrule(l){2-3}  \cmidrule(l){4-5} \cmidrule(l){6-7}
1950.0 &  15\,GHz   &  23\,GHz  &   15\,GHz   &  23\,GHz & 15\,GHz & 23\,GHz\\
\noalign{\smallskip}
\hline \noalign{\smallskip}
0059+581& 1.49 &   6.22  &  $-$0.44 &  $-$0.57 & 0.2 & 0.56\\

0149+218& 1.89 &  0.72   & 0.15      & $-$0.22 & 0.2 & 0.54\\

0241+622& 0.77  &   0.81  & 0.30     & 0.27 & 0.19 & 0.53\\

0528+134& 0.53 &   0.84  & 0.65 &  $-$0.29 & 0.21 & 0.57\\

0748+126& 2.64 &  2.08   & 0.51 & $-$0.37 & 0.2 & 0.55\\

1127$-$145&  1.9 &  2.13   & 0.14 & $-$0.49 & 0.19 & 0.52\\

1243$-$072$^\star$& 4.34 &   1.49  & 0.39 &  0.72 & 0.2 & 0.66\\

1546+027$^\star$&  3.43 &  4.00   & 0.27 &  0.89 & 0.19 & 0.53\\

2136+141& 1.6 &  2.25   &  0.22 & 1.5 & 0.22 & 0.6\\ \noalign{\smallskip} \hline

\end{tabular}
\end{adjustbox}
\end{center}
\tablefoot{Integrated values of polarization fractions for all science targets. \tablefoottext{a}{Fractional linear polarization.}
\tablefoottext{b}{Fractional circular polarization.} \tablefoottext{c}{Gain uncertainty.} \tablefoottext{$\star$}{Excluded from interpretation  (see App.~\ref{app:ex}).}}
\label{tab:pol}
\end{table}

\begin{figure*}[h]
    \centering
    \includegraphics[trim={0cm 0cm 0cm 4cm},clip,width=\textwidth]{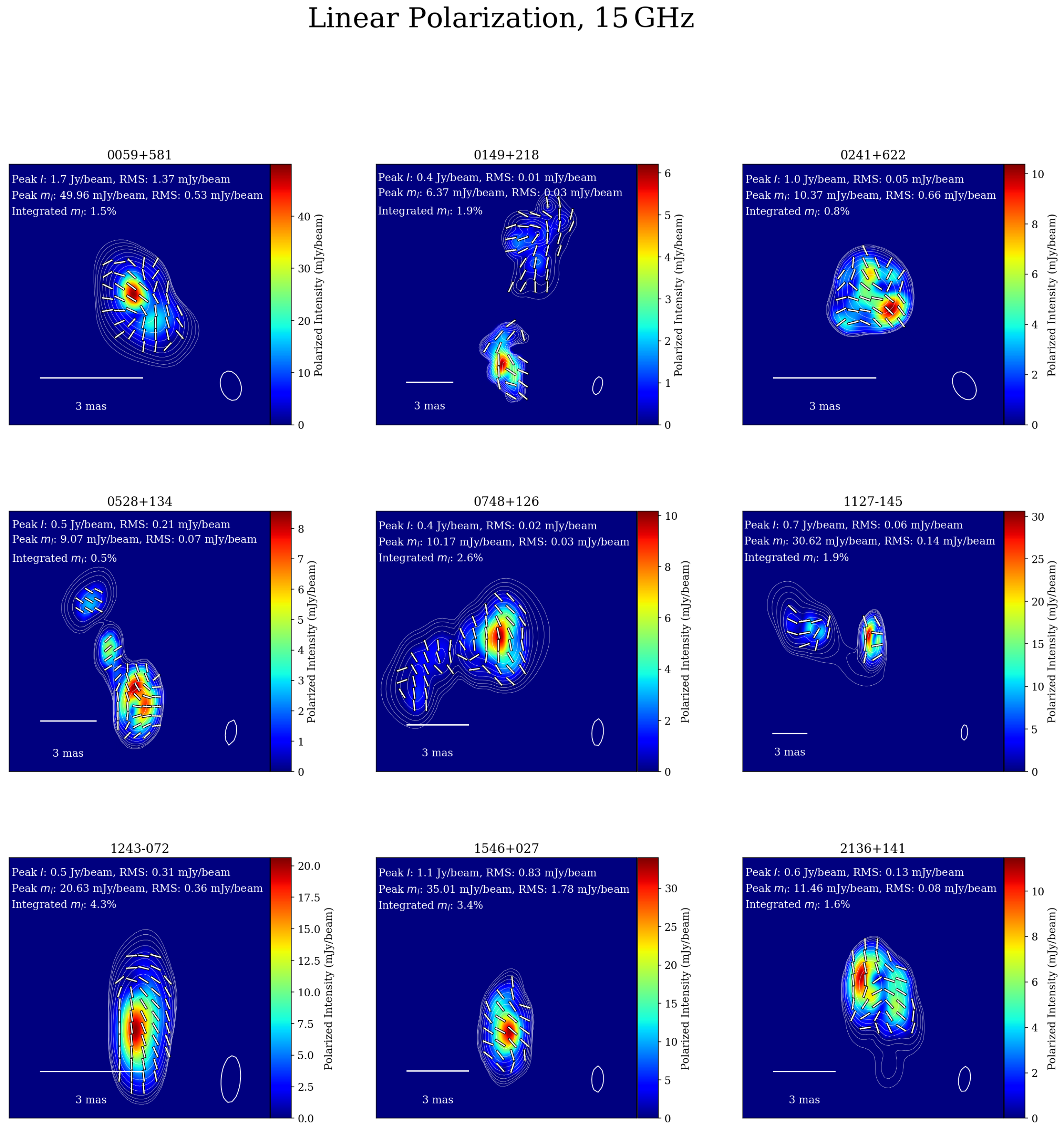}
    \caption{Reconstruction of data from 
    6-7 Jan. 2022 for nine target sources: 0059+581, 0149+218, 0241+622, 0528+134, 0748+126, 1127$-$145, 1243-072, 1546+027, and 2136+141. The figure shows maps for each source in linear polarization at 15\,GHz (colorcoding: blue/red low/high). The total intensity is indicated in white contours on each map. The contours are the $[0.1\%,0.2\%,0.4\%,.., 25.6\%,51.2\%]$ levels of the peak brightness. The orientation of the linear electric vector position angle is plotted as white ticks, the total polarized intensity by the colormap. An individual convolution beam (natural weighting) is shown in the lower right corner of each map. The field of view is shown by the scale with $3\,$mas in the lower left.}
    \label{fig:LP15}
\end{figure*}

\begin{figure*}
    \centering
    \includegraphics[trim={0cm 0cm 0cm 4cm},clip,width=\textwidth]{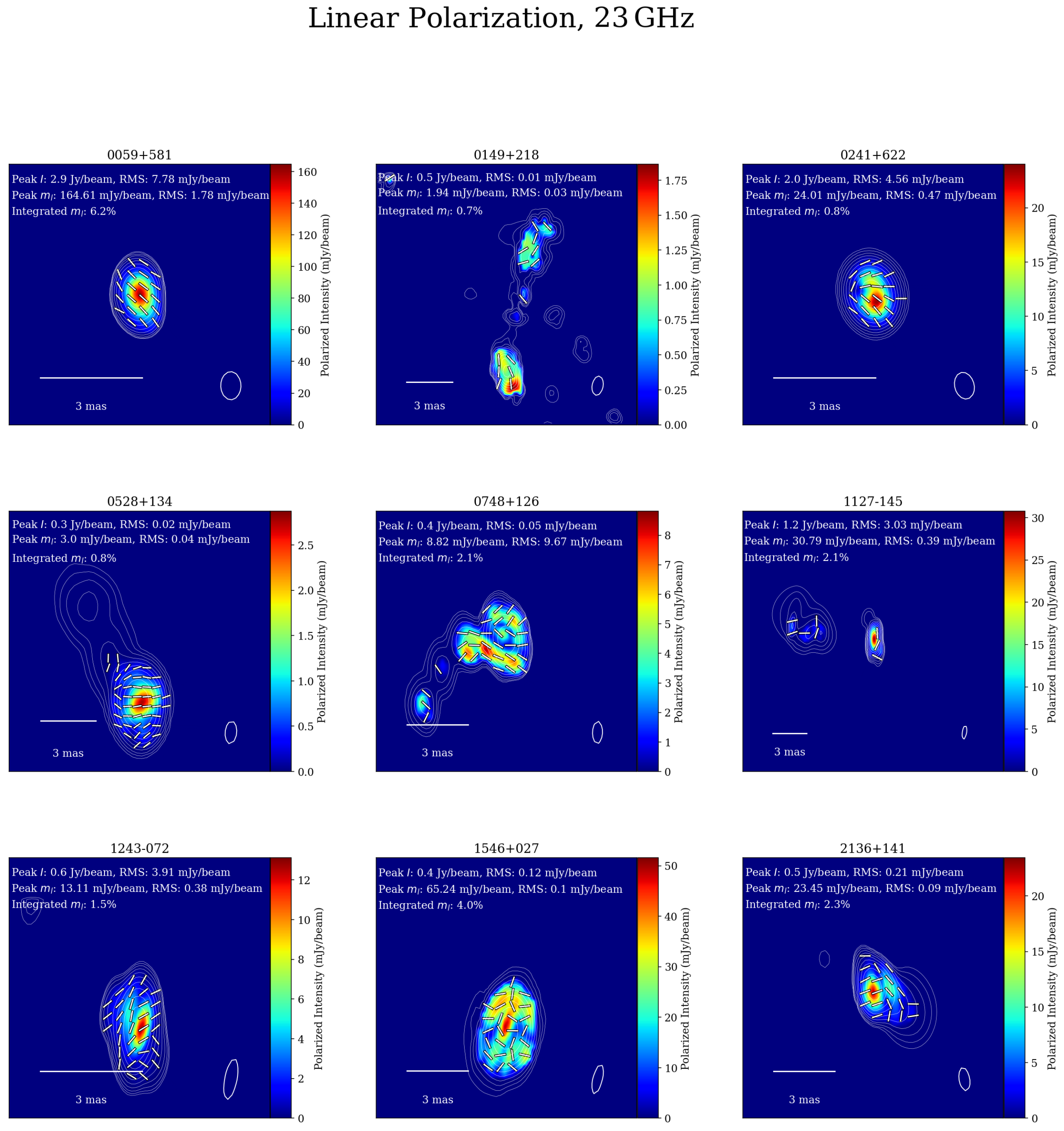}
    \caption{The same as Fig.~\ref{fig:LP15} but at 23\,GHz observational frequency.}
    \label{fig:LP23}
\end{figure*}

\section{Results}\label{sec:Results}
Full-Stokes maps for all target sources are displayed in Figs.~\ref{fig:LP15}--~\ref{fig:CP23}, each convolved with a Gaussian beam. The beams are shown in the lower right corner of each map, while their sizes can be found in Table~\ref{tab:source_list}.  
Our records of total intensity and polarization measurements in both 15\,GHz and 23\,GHz are summarized in Table~\ref{tab:source_list} and the fractional measurements are presented in Table~\ref{tab:pol}. In Table~\ref{tab:source_list}, comprehensive data are provided for total intensity (Stokes $I$), linear polarization (LP, $\sqrt{Q^2+U^2}$), and circular polarization (CP, Stokes $V$) at the peak of each map. 
The peak total intensity in Jansky (Jy) for all nine observed blazar sources range between $0.36^{\pm 10^{-5}}\leq I\leq 1.7^{\pm1.4\times 10^{-3}}\,\mathrm{Jy/beam}$ for 15\,GHz observational data and $0.33^{\pm2\times10^{-5}} \leq I \leq 2.9^{\pm7.8\times10^{-3}}\,\mathrm{Jy/beam}$ at 23\,GHz (see upper left corner in each map in, e.g., Fig.~\ref{fig:LP15} and Fig.~\ref{fig:LP23})\footnote{It is important to note here the uncertainty of the total flux derived from the CLEAN imaging process and frozen into the calibration process.}
The total intensity peak value remains consistent at both 15\,GHz and 23\,GHz wavelengths. 
For a comparison to our values presented in Table~\ref{tab:source_list}, see the \hyperlink{https://www.cv.nrao.edu/MOJAVE/allsources.html}{MOJAVE data archive}.
Evident jet structures are observed in 0149$+$218, characterized by a prominent bright spot to the north. An extended jet structure is further observed in 0528$+$134, 0748$+$126, and 1127$-$145. 
Notably, at both frequencies a counter-jet is newly observed in 0149+218 to the south and in 2136+141 to the east at 23\,GHz, marking the first occurrence of this phenomenon in these sources that is mentioned in a publication. The Quasar 0149+218 has a jet speed of $(14.29\pm0.76)\,{\rm c}$~\citep{Lister2019} -- a strong argument against the visibility of a counter-jet. However, the Bordeaux VLBI Image Database shows a very similar structure to the map displayed in their database~\citep{website}.

The core component of the Quasar 0149+218 shows a cross-like feature usually not seen in archived CLEAN images. To a smaller degree, we see the same structure in the Quasar 0059+581. Since also the jet in the quasar 0149+218 shows more fine-structure (components) than in usual images derived with CLEAN (although that may be explained when \texttt{DoG-HiT} is more sensitive to small scale structure), we add some discussion of this phenomenon here. Note that with \texttt{DoG-HiT} we apply a non-linear minimization scheme fitting the amplitudes and closure quantities, rather than an inverse modelling scheme working with the visibilities. In Fig. \ref{fig:0149_I} we show the reconstructions in total intensity with CLEAN, and \texttt{DoG-HiT}. Moreover, we show the reconstruction achieved by \texttt{DoG-HiT} when only fitting closure phases and closure amplitudes, and the reconstruction achieved when fitting only closure quantities with a LASSO approach. LASSO assumes that the model is sparsely represented in the pixel domain, i.e., imprints the same assumption that goes into CLEAN (although it still differs significantly from CLEAN as a forward modeling approach). We see that the LASSO reconstruction is relatively similar to the \texttt{DoG-HiT} reconstruction using the same data properties, with the difference that latter one filtered out the over-resolved core. That may indicate that the cross-like structure, as well as the spurious components in the jet are not introduced by the wavelet approach, but may be caused by the data fidelity terms.

We see core components to the East and to the West for all reconstructions that utilize closure quantities, in contrast to CLEAN which works with the visibilities. When blurred to the CLEAN resolution the structure (especially in the jet) resembles the CLEAN one, which is natural. The features are more prominent when we factor in amplitudes, rather than only closure quantities. There may be two possible interpretations here. First, it could be that there are non-station based errors that get more highlighted and projected to the images when we rely on the closure quantities only, or the statistics is just smaller introducing spurious artifacts (but then the potential artifacts should be less prominent when using amplitudes additionally). Second, it may be that the imaging with \texttt{DoG-HiT} (which is robust against any phase corruptions) may give rise to more fine-structures that are iteratively removed from the CLEAN reconstructions, e.g., when the phases in the early iterations are potentially too much aligned around the dominant core.

Ultimately, both solutions fit the data. In this manuscript, we go on with the solution derived by \texttt{DoG-HiT} (derived from the amplitudes and closure quantities) due to consistency with the other sources, and since a manual inspection of the amplitudes did not look suspicious to us. For reference, we show in Fig. \ref{fig:0149_pol} the polarization results that we obtain when using only closures. The linear polarizations match very well. However, we observe quite significant changes in circular polarization, although not changing our interpretation.

\subsection{Polarized intensity maps}
Figure~\ref{fig:LP15} shows the linear polarization at 15\,GHz overplotted with both the 
contours in total intensity and EVPAs as white ticks. The same plotting scheme is 
illustrated in Fig.~\ref{fig:LP23} for 23\,GHz. The circular polarization maps at 
15\,GHz and 23\,GHz are presented in Fig.~\ref{fig:CP15} and Fig.~\ref{fig:CP23}, 
respectively. 
In all observed blazars the peak of the Stokes $I$ map corresponds to the compact and 
resolved VLBI core. 
The VLBI core is consistently linearly polarized.
Minor discrepancies arise between 
circular polarization and the peak in total intensity specifically within the VLBI core 
(with no impact on the extended jet structure). This discrepancy is likely attributed to 
the phase calibration process (detailed in Sect.~\ref{sec:calibration}), which could potentially 
introduce a minor positional shift between the location of circular polarization and the 
Stokes $I$ peak.  The overall 
values for the fractional linear and fractional circular polarization [\%] are in a 
range of $0.53^{\pm0.21}\leq m_\mathrm{l} \leq 4.34^{\pm 0.4}$ and $0.14^{\pm0.19}\leq |m_\mathrm{c}| \leq 0.65^{\pm0.21}$ at 15\,GHz, 
and $0.72^{\pm 0.2} \leq m_\mathrm{l} \leq 6.22^{\pm 0.2}$ and $0.22^{\pm 0.2}\leq |m_\mathrm{c}| \leq 1.5^{\pm 0.22}$ at 23\,GHz. 
For reference and details see Table~\ref{tab:pol}.
\looseness=-1

\subsubsection{Polarized structure}\label{sec:Polstructure}
At 23\,GHz, the peak of \emph{linear polarization} in the target source 0748+126 is displaced from the VLBI core with respect to the total intensity $I$ in the linear polarization map (see Fig.~\ref{fig:LP15}). This 
behavior for the linearly polarized structure is observed in 0149+218 (23\,GHz), 0241+622 (mainly 
15\,GHz), 0528+134 (15\,GHz), 0748+126 (23\,GHz) and 2136+141 (15\,GHz).

As for the \emph{circular polarization}, there is a dominant pattern of a two-sign circular polarization (CP) structure at 23\,GHz (Fig.~\ref{fig:CP23}), which contrasts with a positive trend at 15\,GHz (Fig.~\ref{fig:CP15}). 
This dominant sign alteration, however, exhibits hints in the sources 0241+622 and 0748+126.
In the case of the blazars 0059+581 and 2136+141, a consistent and predominantly one-signed radio core is observed across both 15\,GHz and 23\,GHz. Most of the sources show an alignment between the circularly polarized intensity peak and the total intensity structure. This suggests the mechanism producing CP works effectively at the jet's base, near the $\tau = 1$ surface, where $\tau$ represents optical depth~\citep{homan_wardle_1999}. It's crucial to remember that the `core' as the optically thick jet base~\citep{Blandford1979} is theoretical and matches the observed "core" only in high-resolution observations~\citep{Vitri}. Generally, the position of the V peak is influenced by the magnetic field strength and electron density, indicating an expected strong CP signal. One scenario could be that the circular polarization peak emerging slightly before the $I$ peak signals the upcoming emergence of a new VLBI component~\citep{Vitri}.

In general, \emph{the fractional CP}
tends to be higher at 23\,GHz compared to 15\,GHz (see 
Table~\ref{tab:source_list}). 
We are confident that these findings are not the result of observational artifacts. When analyzed collectively or individually, the 23\,GHz $|\bar{m}_\mathrm{c}|$ values exceed those at 15\,GHz, as shown in Tab.~\ref{tab:pol}. The observation that $m_\mathrm{c}$ increases with frequency in several sources, consistently showing higher values at 23\,GHz than at 15\,GHz frequency, aligns with~\citep{WardleHoman2003}. They suggested that CP, whether from synchrotron radiation or Faraday conversion in a \citet{Blandford1979} jet, could show an inverted CP spectrum, $m_\mathrm{c} \propto \nu$. This implies our measurements might reflect the intrinsic inhomogeneity of the jet affecting the observed CP and its spectrum. Moreover, the complexity increases as core CP measurements could blend CP contributions from various regions within the core and the innermost jet, a phenomenon directly observed in 3C\,84~\citep{Homan2004}.

The sources 0059+581, 0149+218, 0748+126, and 1127$-$145 show an elongated \emph{EVPA structure} along the jet direction at 15\,GHz (see Fig.~\ref{fig:LP15}). For 0528+134 
(and 1127$-$145 to some extent) we observe the EVPAs to be perpendicular to the jet 
direction at both frequencies (see Fig.~\ref{fig:LP15} and Fig.~\ref{fig:LP23}). 

\begin{figure*}[h!]
    \centering
    \includegraphics[clip,width=0.465\textwidth]{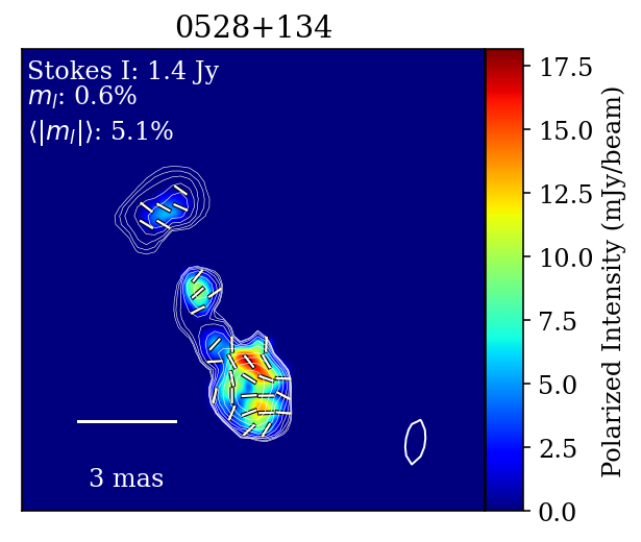}
    \includegraphics[clip,width=0.465\textwidth]{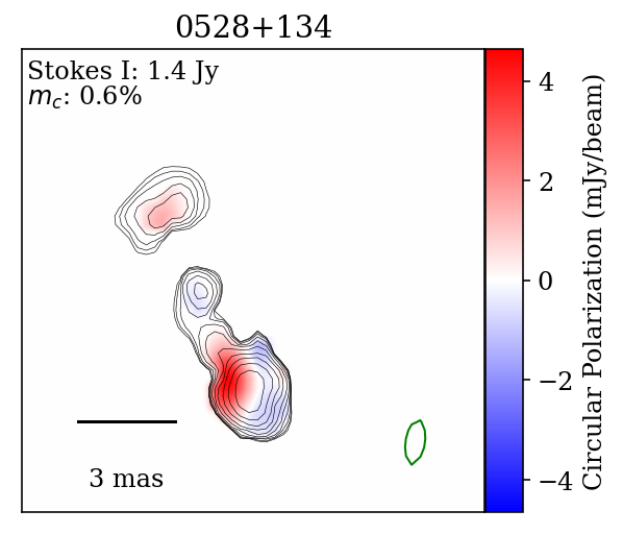}
    \caption{Reconstruction of data from 6-7 Jan.
    2022 for the quasar 0528+134. The images show the resolved maps of: Linear polarization at 15\,GHz (color coding: blue/red low/high) in the left panel. The total intensity is indicated in white contours. The linear electric vector position angle is plotted as white ticks. Right: circular polarization (colorcoding: blue/red negative/positive). The total intensity is indicated in white contours.}
    \label{fig:superres}
\end{figure*}

\subsubsection{Amplitude of polarization}\label{sec:Polvalues}
We detect circular polarization (and linear polarization) in all nine target sources.
The distribution of the peak values of linear 
and circular polarization is consistent throughout all sources; specifically, the amplitude 
in linear polarization (Stokes $Q$ and $U$) is consistently higher than the circular 
polarization amplitude (Stokes $V$). The ratio of the fractional values of linear 
polarization $m_\mathrm{l}$ to circular polarization $m_\mathrm{c}$ vary to a greater 
extent.\footnote{from this point on we will refer to LP and CP as fractional values.} 
0059+581, 0149+218, 0748+126, 1127-145, 1243$-$072, (1546+027), and 2136+141, have significantly higher fractional linear polarization $m_\mathrm{l}$ values compared to fractional circular polarization $m_\mathrm{c}$ at 15\,GHz and 23\,GHz. 
Overall, there is no clear trend for either frequency nor in the fractional LP 
when comparing 15\,GHz and 23\,GHz.
\looseness=1

\subsection{Superresolution}
\texttt{DoG-HiT} achieves superresolution by combining the significant advantages of regularized maximum likelihood reconstructions, which super-resolve structures, with the CLEAN method, which provides high dynamic range sensitivity to extended structures. For CLEAN, the model that is fitted to the visibilities is a list of $\delta$-components which is an unphysical representation of the on-sky image. Therefore, we need to convolve the model with the clean beam for a proper representation of the image. Regularized maximum likelihood reconstructions, multiscalar CLEAN variants, or Bayesian approaches which fit smoothed, spatially correlated functions to the visibilities may not need this final convolution. The fitted model itself can be already seen as a proper representation of the sky brigtness distribution, a potential way towards super-resolution under the limitation that the algorithms may pick up noisy small-scale features. We refer the reader to \citet{Honma_2014} for an illustrative discussion of this point, and to \citet{Mueller2023a} for a demonstration in a CLEAN-based environment. Remind from Sec. \ref{sec:calibration} that this fact was one of the main motivations for the imaging and calibration strategy applied in this manuscript.

Without applying a Gaussian beam convolution (in contrast to Figs.~\ref{fig:LP15}-\ref{fig:CP23}), we obtained a superresolved image (see Fig.~\ref{fig:superres}) that fits the observed visibilities similarly as the list of $\delta$-components in CLEAN. We chose the blazar 0528+134 at 15\,GHz as an example, since the source shows the richest structural features along the jet (see Fig.~\ref{fig:CP15}).
The blurred map (fourth panel in Fig.~\ref{fig:CP15}) matches the reconstructed 15\,GHz map available in the MOJAVE data base quite well, compare also Fig. \ref{fig:15clean} and Fig. \ref{fig:15doghit}. The difference that we achieve by illustrating the 
super-resolved jet in Fig. \ref{fig:superres} is that it shows two distinct features further upstream, whereas the 
MOJAVE image would resolve a single upstream (polarized) feature. 
This suggests that the improvements made to our pipeline allow
us to resolve distinct features at lower frequencies. 

In this manuscript, we mainly report the net polarization rather than the total polarization fraction due to its advantages in the scientific interpretation. Both quantities are expected to match for marginally resolved sources with a preferred EVPA pattern, but differ the more resolved and scrambled the EVPA pattern appears, as is the case here. To ease comparison with other observations, we also report the total polarization fraction in Fig. \ref{fig:superres} which is significantly higher than the net polarization.

Past analyses of the trajectories of individual moving features show that the bend is more 
pronounced closer to the core and gradually straightens out farther away \citep{Britzen99}. 
Curvature on small angular scales has been observed in numerous other sources, suggesting 
that it is a common phenomenon in AGN (for example, \cite{zensus1995} for 3C\,345 and 
\cite{wagner1995} for PKS\,0420$-$014). A possible explanation for the observed 
"wiggling" of 
the core region is the ejection of moving features along different, but straight 
trajectories, resembling ballistic motion \citep[e.g.,][]{jorstad:2017}.

The EVPAs are oriented perpendicular to the jet's direction, matching an underlying poloidal 
magnetic field morphology~\citep{Kramer2021} and being consistent with the 43\,GHz 
observations presented in~\cite{Palma}. 


\section{Discussion}\label{sec:discussion}
It is imperative to highlight \emph{improvements in the imaging and calibration procedure} that
could potentially account for the cases of high circular polarization. Before \citep{homan_wardle_1999} and following, it was not possible to detect reliable resolved CP in AGN. The usual level of detectable CP was around 0.1\,\%~\cite[see early studies in, e.g.,][]{Weiler, Komesaroff}. Similar to~\citep{homan_wardle_1999} and following years we can not conclude on any mechanisms for the production of CP, however, with our methods and the application and comparison of their method in our dataset, we can interpret favored magnetic field configurations. 

In particular, the accuracy of Stokes $V$ strongly hinges on meticulous calibration. To be more 
precise, our self-calibration is anchored in a physical model, as opposed to relying on a 
potentially non-physical model of CLEAN (delta) components. This adjustment results in potentially more
accurate gains. 
 
Fig.~\ref{fig:rrll} shows the amount of detected circular polarized emission over the $uv$-
distance. It further stresses that the observed circular polarization is primarily detected 
on shorter baselines. The observed data (in blue) 
align well with the reconstructed model (in red) within the noise budget. There is no significant trend over the 
distance, nor are there any major outliers within our calibration method. 

While the closure phases for RR and LL remain resilient to variations in antenna gains, they could potentially be affected by instrumental polarimetric leakage. However, the effect of uncertainties in $D$-terms is significantly more pronounced in cross-hand visibilities compared to parallel-hand visibilities~\citep{Smirnov_2011}. Consequently, this implies that instrumental polarization exerts a comparatively lesser effect on Stokes $V$ in comparison to its impact on Stokes $Q$ and $U$. For our data we hence stress again that the Stokes $V$ signal is robust and reliable. The rotation of the EVPAs could slightly change depending on the solution of the leakage terms.

We need to comment however that for the two sources 1243-072 and 1546+027 the recovered CP structure depends strongly on the calibration method and is most likely driven by gains. The interpretation of the recovered maps is based on the morphologies in LP and CP across the two frequencies. We need to highlight however as a limiting factor that the CP signal at 23\,GHz suffers from large gain-driven uncertainties with only a few significant detections.

The presence of substantial \emph{fractional circular polarization}, reaching up to $|\bar{m}_\mathrm{c}|=1.5\,$\%
in our dataset at both frequencies, exceeds previously observed circular polarization 
rates documented by, e.g.,~\cite{Homan2001, homan2006mojave}\footnote{Notably, they 
identified circular polarization exceeding 0.3\,\% in merely two out of 36 sources.} using 
the VLBA at 5\,GHz or by MOJAVE observations with the VLBA at 15\,GHz~\citep[][where CP was 
around 0.3 percent]{homan2006mojave}. 
We agree with their conclusion that the overall circular polarization value tends to 
increase at higher frequencies~\citep[see also][]{Vitri}. 

Although the observed circular polarization could theoretically originate from intrinsic CP via synchrotron radiation, certain observed $m_\mathrm{c}$ values are too elevated to justify by this means alone, necessitating either unrealistically strong or exceptionally orderly magnetic fields.
In some instances, $m_\mathrm{c}$ values exceeding 1\,\% have been noted~\citep{homan2006mojave,Homan2004,Vitri}.
Another case of remarkably high fractional circular 
polarization was detected in 3C\,84, exceeding $m_\mathrm{c}=3\,$\%~\citep{Homan2004}.
Therefore, it's likely that an alternative mechanism is at work in certain scenarios. This mechanism is thought to be Faraday conversion, transforming linear polarization into circular polarization within a magnetized plasma~\citep{Gabuzda2021, OSullivan2009}.
\\\\
Most of our sources exhibit stable degrees of \emph{fractional linear polarization} over the last 
years (monitored by the MOJAVE program). Eight out of nine sources remain 
consistent in the peak brightness of the linear polarization $P$, the polarization angle, and the degree of polarization $\bar{m}_\mathrm{l}$ compared to the archival data collected over several years to 
decades in the \hyperlink{https://www.cv.nrao.edu/MOJAVE/allsources.html}{MOJAVE data archive}. The only exception is 0149+218 which shows a dimmer peak in our data. $0241+622$ has no previous information on EVPA in the MOJAVE database, the values of our polarization study, however, are comparable with observations performed in 2007. The offset between the total intensity and linearly polarized intensity seen in 0528+134 is consistent with polarized maps in the MOJAVE dataset. For the blazars 0748+126 and 1127$-$145, our study results in a lower peak brightness in Stokes $I$.

In the cases of 0059+581 and 2136+141, the jet structure appears to be oriented further to the south (rather than west) in 15\,GHz compared to its appearance in 2013 and 2023, respectively (see 0059+581 and 2136+141 in the \hyperlink{https://www.cv.nrao.edu/MOJAVE/sourcepages}{MOJAVE data archive}).
\\\\
\cite{Kramer2021, Kramer24} provided an analysis on how the magnetic field morphology affects the 
polarized synchrotron emission within the recollimation shock (VLBI core) and the extended 
jet structure. They found a centrally highlighted VLBI core and a single sign in CP for a 
purely poloidal magnetic field. In contrast to that, a purely toroidal magnetic field would 
result in an edge-brightened jet, a bi-modal EVPA pattern (where the EVPAs tend to 
align with the jet's direction of motion within the spine), and a two-signed CP structure 
within both the VLBI core and the jet. A helical magnetic field combines these 
morphological characteristics. 

In the context of theoretical forecasts, a tendency for a favored \emph{magnetic field orientation} becomes 
evident within the extended jet emission of, for instance, 0241+622: its polarized structure suggests an 
underlying toroidal magnetic field structure. This conclusion is supported by two 
observations at 15\,GHz, namely, a bi-modal pattern of EVPA (see Fig.~\ref{fig:LP23}) and 
the presence of two signs in CP (see Fig.~\ref{fig:CP23}). 

The \emph{jet structure} in 0528+134 observed at 15\,GHz in superresolution, see 
Fig.~\ref{fig:superres}, exhibits characteristics that align with a helical or rather 
poloidal magnetic nature. 

The blazar sources 0059+581, 0748+126, and 1127$-$145 are in agreement with a helical 
magnetic field morphology. The EVPA tend to follow the jets motion (even when bending), 
however, the central peak of the linear polarization is slightly off-center from the total 
intensity peak (see, e.g., Fig.~\ref{fig:LP15}). 

For the \emph{radio core}, the configuration in 0528+134, 0748+126, 1127$-$145,
and 2136+141 presents traits indicative of a helical or toroidal magnetic field structure, 
supported by a double sign in CP in either frequency (except for 1546+027 which is showing the bimodality only at 15 GHz). We note however, that the bimodality observed for 1243$-$072 and 1546+027 are not robust against changing the calibration strategy. Consequently, the interpretation of a toroidal magnetic field configuration may be an over-interpretation for this source. Asymmetric bimodal structures may be caused by phase errors \citep{homan2006mojave}. However, we trust the found source structures for the other sources due to their robustness against multiple calibration techniques (except 1243$-$072 and 1546+027), and against a final self-calibration as suggested in \citet{homan2006mojave}. 
An offset between the linear polarized intensity peak and the total intensity peak is visible in, for instance, 0241+622 and 2136+141, 
which is favored in a helical treatment of the magnetic field~\citep{Gabuzda2018,Gabuzda2021}.  This aligns with synthetic polarized emission maps presented in \cite{Kramer2021}. When focusing on the radio core of the sources 0059+581 or 0748+126 an emphasis can be placed on a purely poloidal magnetic field structure. This is summarized in Table~\ref{tab:magfield}.
\\\\
For quite some time, it has been established that the circular polarization sign within a 
specific AGN tends to remain consistent over extended periods, often spanning years or even 
decades, as highlighted by \cite{homan_wardle_1999}. However, limited information has been 
available concerning the frequency-dependent behavior of the CP sign. \cite{Vitri} findings 
reveal that out of nine AGN where CP was detected at both 15\,GHz and 23\,GHz, eight 
consistently displayed the same sign, with 2251+158 being the exception. Among the six AGN 
where CP was detected at both 23\,GHz and 43\,GHz, four exhibited changes in sign between 
these two frequencies (namely, 0851+202, 1253$-$055, 1510$-$089, and 2251+158). We observe 
the switch in sign for three to five sources as described in Sect.~\ref{sec:Polstructure}. 
The occurrence of alterations in the CP sign as the frequency shifts from 15\,GHz to 
23\,GHz implies the influence of optical depth effects. 
This observation suggests that the sampled regions are likely to be optically thin. Especially, we would like to draw 
attention to the blazar 0241+622 which experiences a complete switch in sign of CP structure from 
15\,GHz to 23\,GHz. The interpretation and findings in CP is supported by the 
offset in EVPA rotation when both frequencies are compared. However, optically thick regions favor the generation of CP with a specific sign influenced by the magnetic field configuration. As the jet transitions to being optically thin, CP may weaken or reverse due to changes in the plasma's characteristics. We are therefore cautious with our interpretation of the magnetic field, however, we do interpret favored magnetic field configurations with a conclusive comparison to synthetic polarized emission maps based on thermal plasma simulations. 

\begin{table}[]
\small
\caption{Preferred magnetic field morphology}
\begin{center}
\begin{adjustbox}{max width=0.5\textwidth}

\begin{tabular}{@{}lccc@{}}
\hline
\hline
\noalign{\smallskip}
IAU & Toroidal\tablefootmark{a} & Poloidal\tablefootmark{b} & Helical\tablefootmark{c} \\
1950.0 &   \\
\noalign{\smallskip}
\hline \noalign{\smallskip}
0059+581 (jet)& (\cmark) & \cmark   \\
0149+218 (jet) & (\cmark) & \cmark \\

0241+622 & & & \cmark  \\

0528+134 (jet)& & (\cmark) & \cmark  \\     

0748+126 (jet)& \cmark & & (\cmark)\\

1127$-$145 (jet)& \cmark &  & (\cmark)\\

1243$-$072$^\star$ (jet)& \cmark &  & (\cmark)\\

1546+027$^\star$& \cmark &\\

2136+141 (jet)& \cmark(\cmark) & &\\ 
\noalign{\smallskip} \hline
\end{tabular}
\end{adjustbox}
\end{center}
\tablefoot{\tablefoottext{a}{Purely toroidal magnetic field configuration.}
\tablefoottext{b}{Purely poloidal magnetic field configuration.}
\tablefoottext{c}{Helical magnetic field configuration.}\tablefoottext{$\star$}{Calibration driven, excluded from interpretation  (see App.~\ref{app:ex}).}}
\label{tab:magfield}
\end{table}

\section{Conclusion}\label{sec:conclusion}
We employ an advanced imaging algorithm using the imaging software \texttt{DoG-HiT}, which potentially shows improvements in the reconstruction of compact structures in an unbiased way. 
Notably, \texttt{DoG-HiT} performs polarized gain calibration using the compact Stokes $V$ 
structure free of biases induced by CLEAN. Our 
calibration method reveals no misleading trends in the right-or left-handed circular 
polarization at different distances, and no major outliers, reinforcing the robustness and 
reliability of the Stokes $V$ signal in our data.

Following~\cite{Gabuzda2018}, we can concur that magnetic fields with a toroidal component 
or with a helical nature are the most plausible explanations for the extended jet 
structures (see Table~\ref{tab:magfield}).
Our analysis leads us to the conclusion that circular polarization serves as a powerful 
tool for elucidating the intrinsic magnetic field morphology of jetted AGN. Through this 
approach, we have been able to associate most of the sources with specific intrinsic 
magnetic field morphology, that is, whether the magnetic field is poloidal or toroidal in 
nature and to identify a favored composition within each.
The findings extracted from our study can be summarized as follows:
\begin{itemize}
    \item[-] The polarized structure, level of polarization, and EVPA orientation over time compared to archival MOJAVE data is very robust (see Fig.~\ref{fig:LP15}).
    
    \item[-] Theoretical predictions favor specific magnetic field orientations within the extended jet structures of the blazar sources \citep{Kramer2021}.
    
    \item[-] The changes in the (dominant) CP sign as the frequency transitions from 15\,GHz to 23\,GHz suggests the influence of optical depth effects (cf.\ Fig.~\ref{fig:CP15} and Fig.~\ref{fig:CP23}).

    \item[-] Two blazar sources show signs of a newly manifested counter-jet structure in total and linearly polarized intensity: 0149+218 to the south (15\,GHz; carrying negative CP) and 2136+141 to the east (23\,GHz; negative sign in CP).

    \item[-] We reconstructed a resolved 15\,GHz map of the blazar 0528+134, which shows superresolved components at lower observational frequency than previously observed.

     \item[-] We avoided relying on a potentially non-physical model of CLEAN components by anchoring our self-calibration to a physical model.
    
    \item[-] Our polarization calibration method departed from previously applied methods that assume $V=0$; instead, we \emph{integrated} the compact Stokes $V$ structure.
\end{itemize}
In order to strengthen the reliability and authenticity of the observed and reconstructed polarized signal, we intend to compare the VLBI data, accounting for leakage terms of each antenna, with the levels of polarized emission obtained through single-dish measurements. To achieve this validation, we intend to utilize data from the G-GAMMA/QUIVER program~\citep[Full-Stokes, multi-frequency radio monitoring of Fermi blazars with the Effelsberg telescope,][]{Fgamma} and the POLAMI program~\citep[Polarimetric Monitoring of AGN at Millimeter Wavelengths with the IRAM 30\,m telescope,][]{Polami}, focusing on quasi-simultaneous observations that align with our study. This approach will also yield essential information about on the absolute electric vector position angle. 

In order to make strong assumptions on a preferred magnetic field, a statistical analysis would be necessary. This could, in principle, be conducted by using the archival MOJAVE data. 
Besides that, to further test and verify the favored magnetic field morphology for the different sources, we plan to compare synthetic transverse intensity profiles with those obtained from the sources presented in this work. In order to stress the conclusions drawn on the optical depth, a future work will focus on the analysis of the spectral index maps obtained between 15\,GHz and 23\,GHz.

\begin{acknowledgements}
J. A. Kramer and H. M\"uller have contributed equally to this work. The authors thank J. D. Livingston for his intense research and fruitful discussion on polarization. We thank J. Kim for his insights on Bayesian imaging techniques on polarized data. We acknowledge I. Myserlis and N. R. MacDonald for their contribution to the observational proposal. We are grateful for D.~Pesce's insights on the calibration steps. 
The analysis was performed with the software MrBeam\footnote{\url{https://github.com/hmuellergoe/mrbeam}} which is partially based on ehtim
\footnote{\url{https://github.com/achael/eht-imaging}}, regpy\footnote{\url{https://github.com/regpy/regpy}} and Wise\footnote{\url{https://github.com/flomertens/wise}}. 
J. A . Kramer is supported for her research by a NASA/ATP project. The LA-UR number is LA-UR-24-22726. H. M\"uller acknowledges financial support through the Jansky fellowship program of the National Radio Astronomy Observatory. J. R\"oder acknowledges financial support from the Severo Ochoa grant CEX2021-001131-S funded by MICIU/AEI/ 10.13039/501100011033.
This research was supported through a PhD grant from the International Max Planck Research School (IMPRS) for Astronomy and Astrophysics at the Universities of Bonn and Cologne. The observations presented are conducted with the Very Long Baseline Array (VLBA) operated by National Radio Astronomy Observatory (NRAO) and correlated at the NRAO correlator in Soccoro. NRAO is a facility of the National Science Foundation operated under cooperative agreement by Associated Universities, Inc.
This work is part of the M2FINDERS project which has received funding from the European Research Council (ERC) under the European Union’s Horizon 2020 Research and Innovation Program (grant agreement No 101018682).

\end{acknowledgements}

\bibliographystyle{aa}
\bibliography{sample.bib}

\begin{appendix}
\section{Validation of results} \label{app:comparison}
\begin{figure}
\centering
\includegraphics[width=\columnwidth]{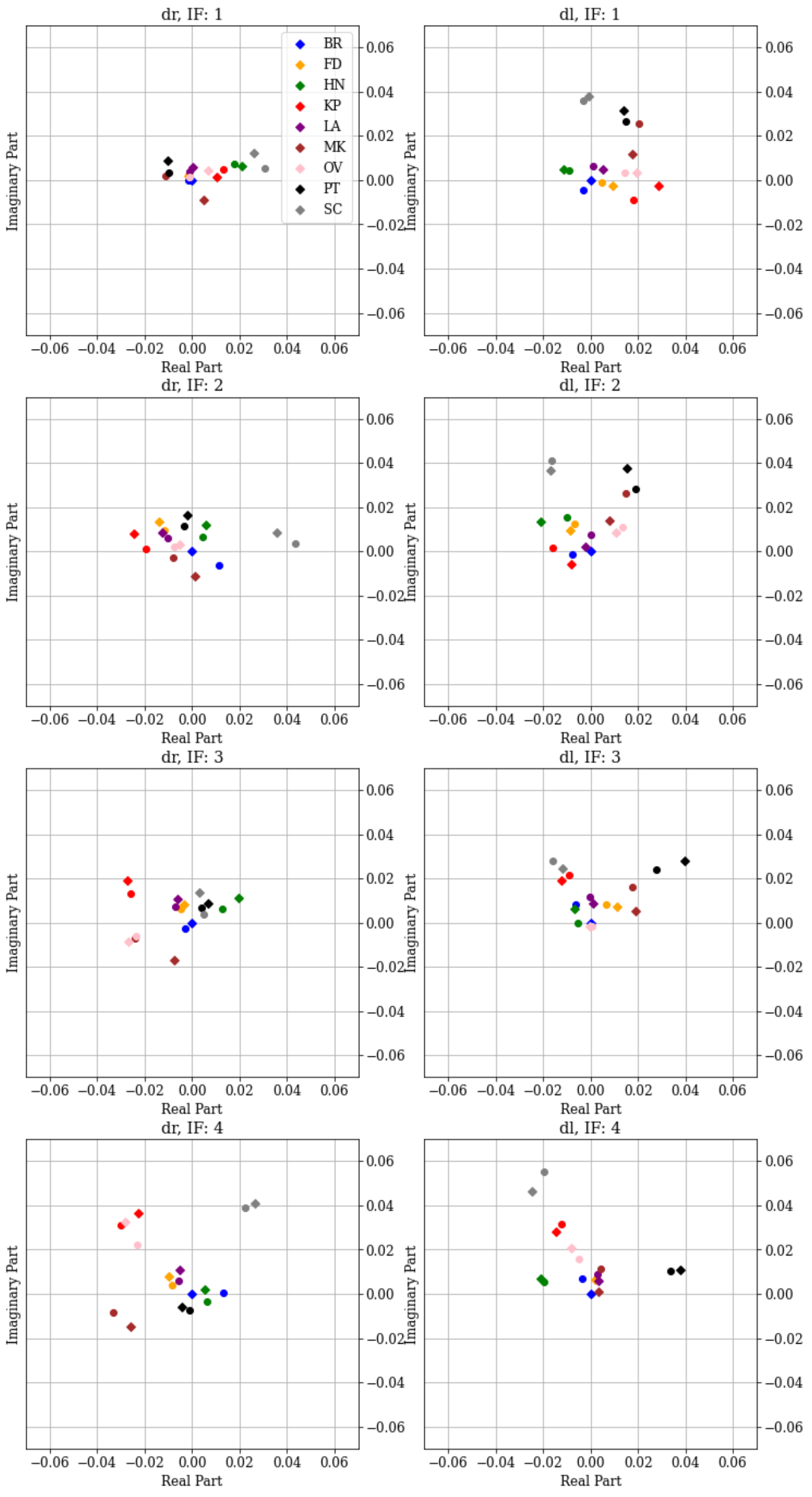}
\caption{$D$-terms: CLEAN (diamond) vs \texttt{DoG-HiT} (circle). Left panels show the right-handed solutions, the right panels the left-handed solutions. Solutions for various IFs are presented in various rows.}
\label{fig:dterms}
\end{figure}

\begin{figure*}
\centering
\includegraphics[clip,width=\textwidth]{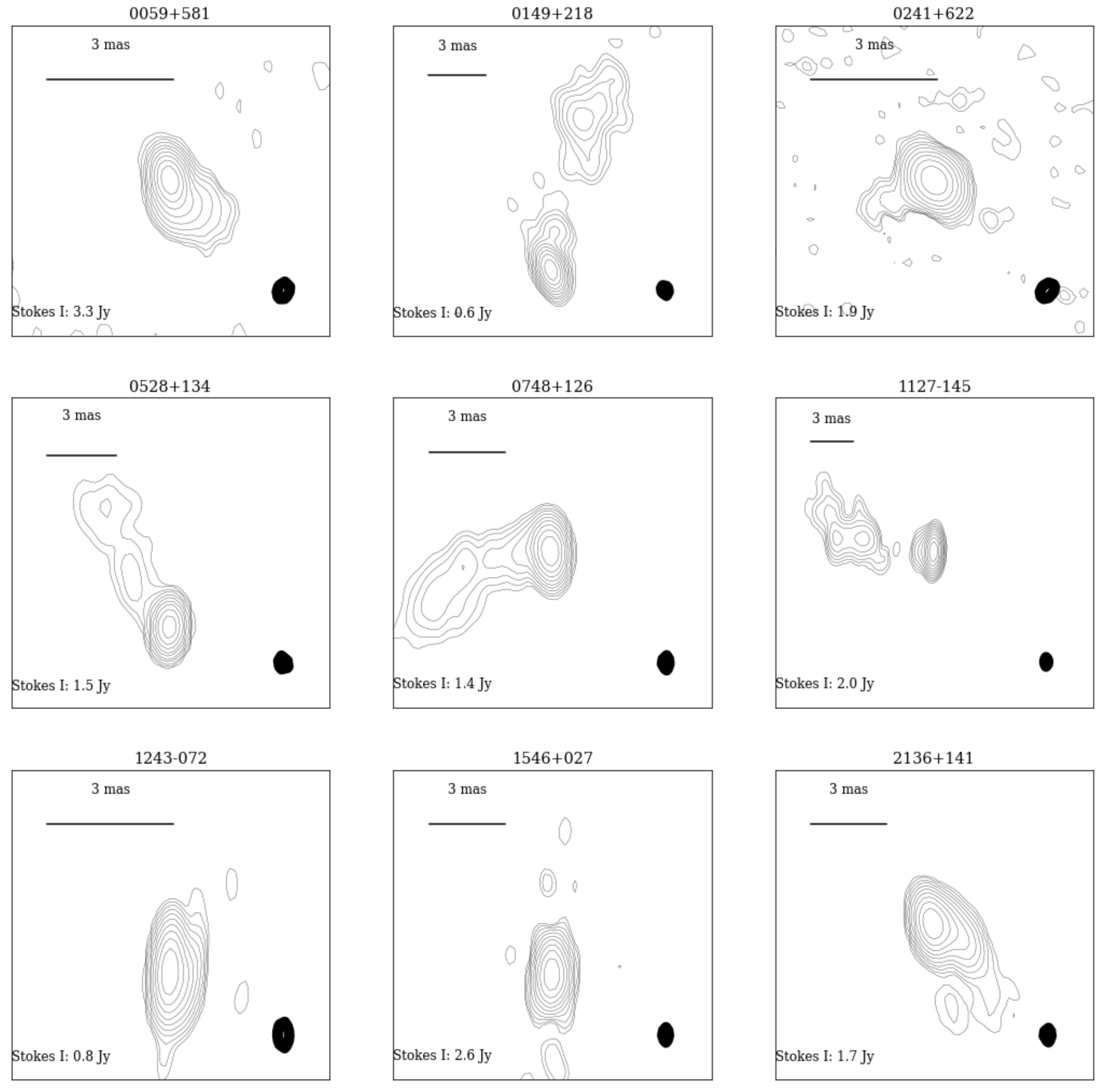}
\caption{Same level plotting comparison between CLEAN and \texttt{DoG-HiT} results: CLEAN. The conotour levels are $[0.1\%,0.2\%,0.4\%,...,51.2\%]$ of the peak brightness emission. The convolution beam has been derived with uniform weighting and is shown in the lower left of the images. The scale of the image is shown in the top left with a bar of $3\,\mathrm{mas}$ in length.}
\label{fig:15clean}
\end{figure*}

\begin{figure*}
\centering
\includegraphics[clip,width=\textwidth]{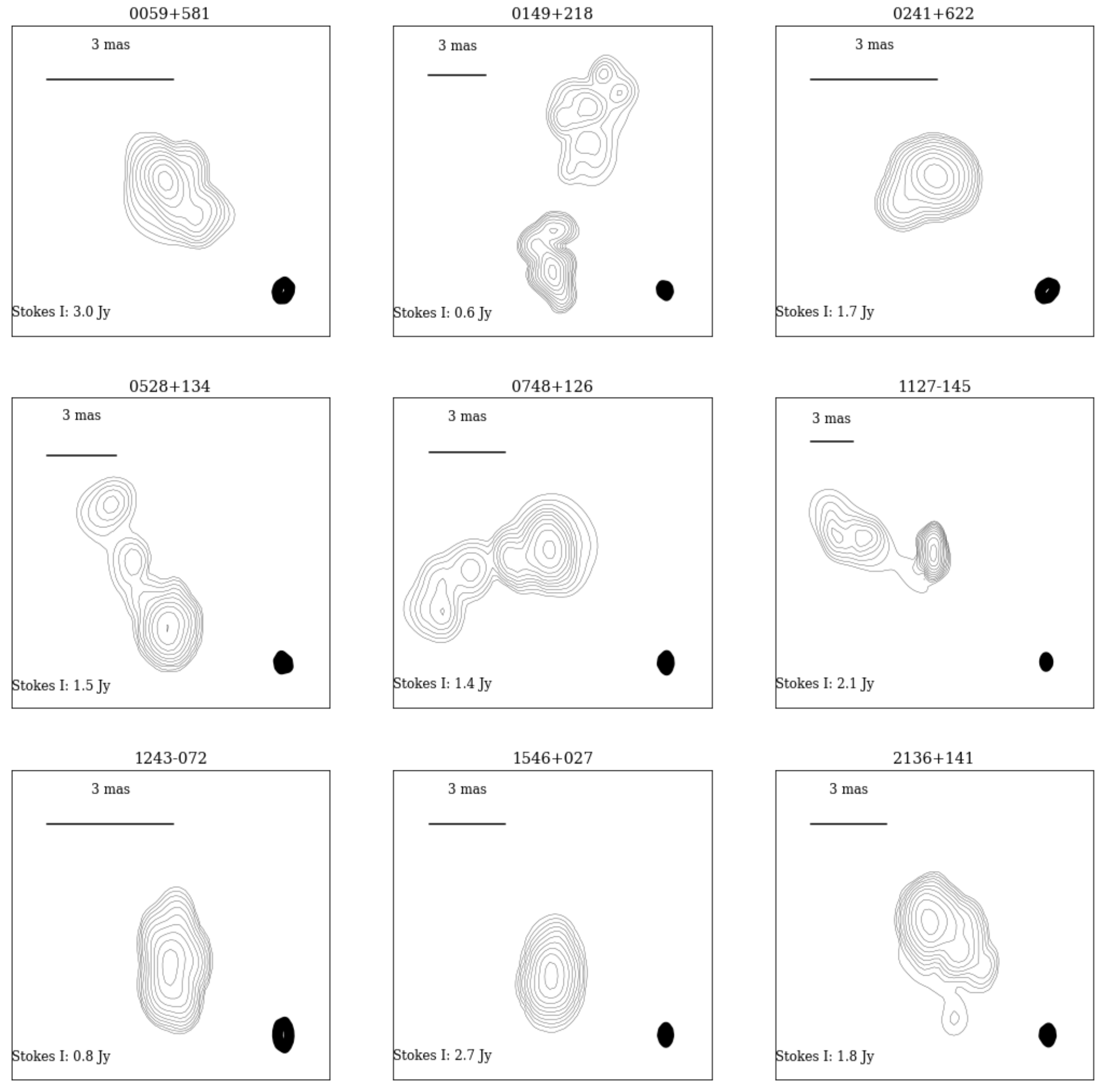}
\caption{Same level plotting comparison between CLEAN and \texttt{DoG-HiT} results: \texttt{DoG-HiT}. To allow for comparisons to the CLEAN images presented in Fig. \ref{fig:15clean}, the reconstructions have been blurred by the uniform weighting beam.}
\label{fig:15doghit}
\end{figure*}

\begin{figure*}
    \centering
    \includegraphics[trim={0cm 0cm 0cm 4cm},clip,width=\textwidth]{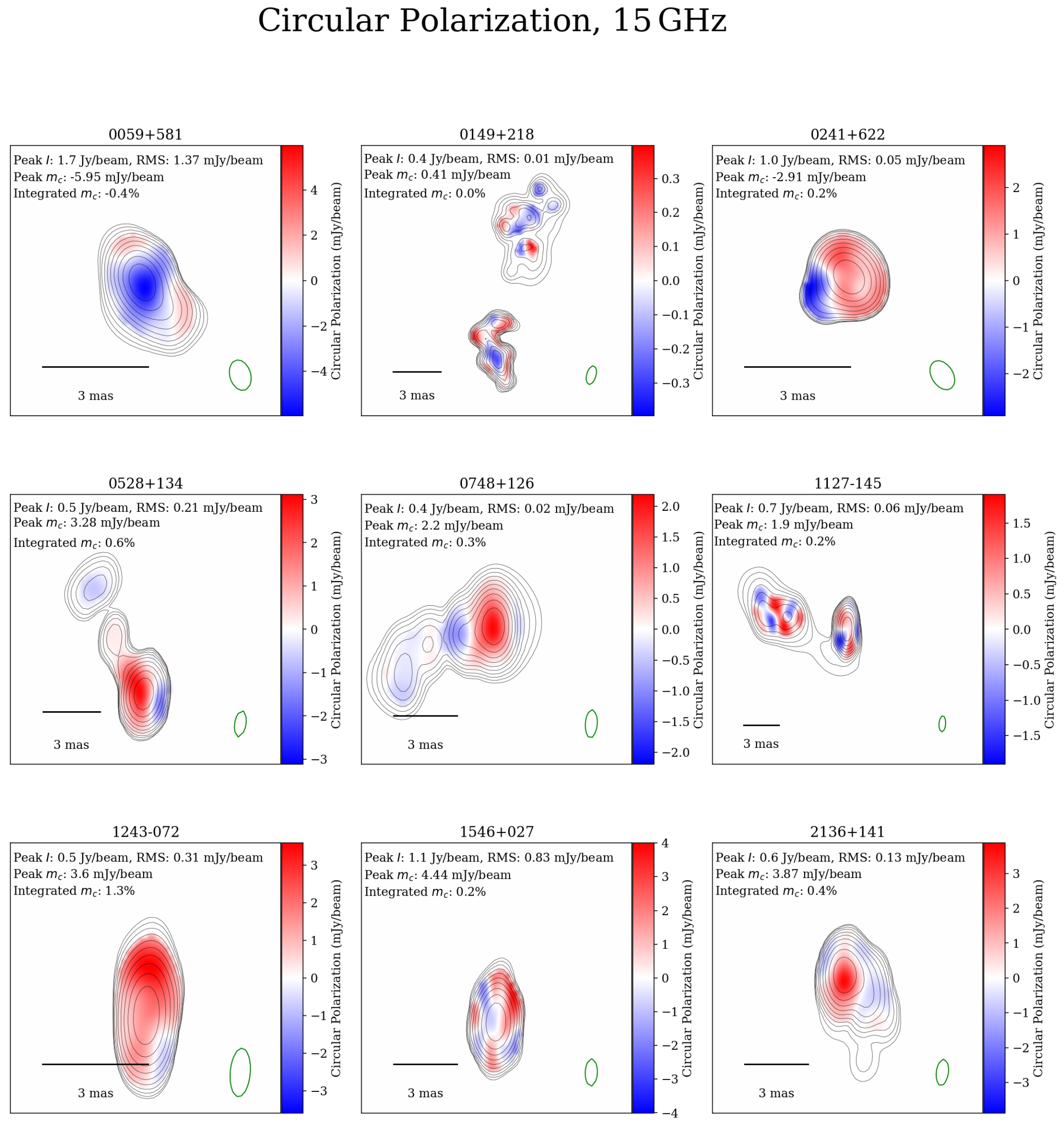}
    \caption{Calibration and imaging based on \cite{Homan2001}. Illustrating the same parameters as in Fig.~\ref{fig:CP15}.}
    \label{fig:CP15homan}
\end{figure*}

The main aspect we would like to highlight in our comparison to, e.g., \cite{Homan2001} are: \emph{i} similar imaging results of linear polarization compared to both MOJAVE and the imaging technique used in \cite{Homan2001} (compare Figs.~\ref{fig:15doghit} \& \ref{fig:15clean}), \emph{ii} the self-calibration and leakage determination are consistent within methods (see further Fig.~\ref{fig:dterms}), and \emph{iii} the imaging technique, i.e., running median, presented in \cite{Homan2001}, and our polarization calibration, focusing on Stokes $V$ (common RL fit), are consistent. See for further verification Fig.~\ref{fig:CP15homan}.

Specifically, in order to verify the results obtained using \texttt{DoG-HiT}, we verify the 15\,GHz total intensity source structure (Fig. \ref{fig:15doghit}) with CLEAN images of all sources (Fig. \ref{fig:15clean}). Apart from the absence of CLEAN artifacts, \texttt{DoG-HiT} mainly reconstructs individual features within the jets more clearly pronounced than they appear in CLEAN. As an example, the extended jet in 0528$+$134 appears diffuse in CLEAN, whereas the \texttt{DoG-HiT} image clearly shows the knotty structure. Similar differences can be seen in 0748+126 and 2136$+$141. \texttt{DoG-HiT} reveals the sub-structure of the inner 5\,mas in 0149$+$218 with a striking clarity in comparison to CLEAN. 

\begin{figure*}
\centering
\begin{subfigure}[b]{0.49\textwidth}
\includegraphics[width=\textwidth]{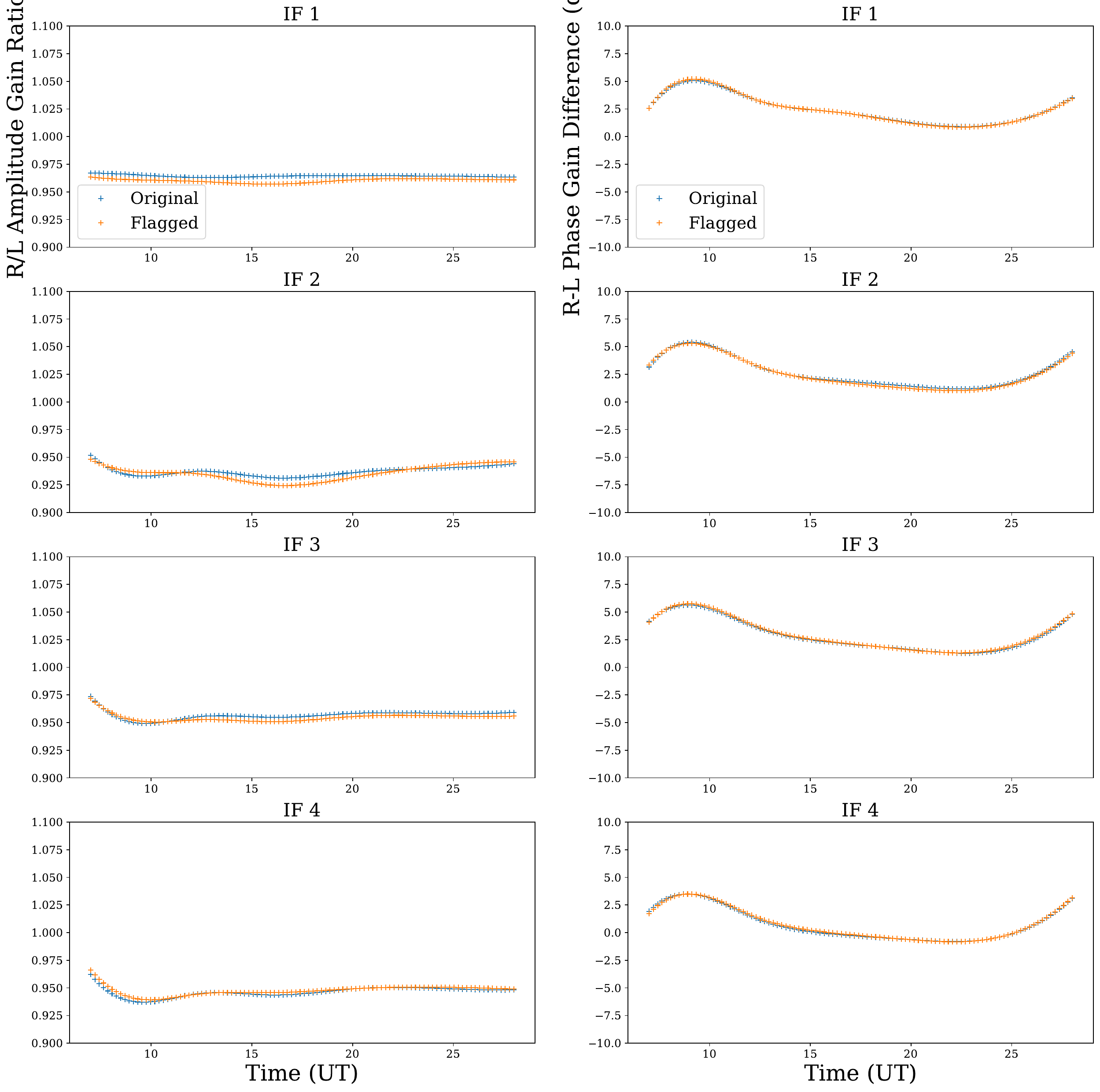}
\caption{Fort Davis}
\end{subfigure}
\begin{subfigure}[b]{0.49\textwidth}
\includegraphics[width=\textwidth]{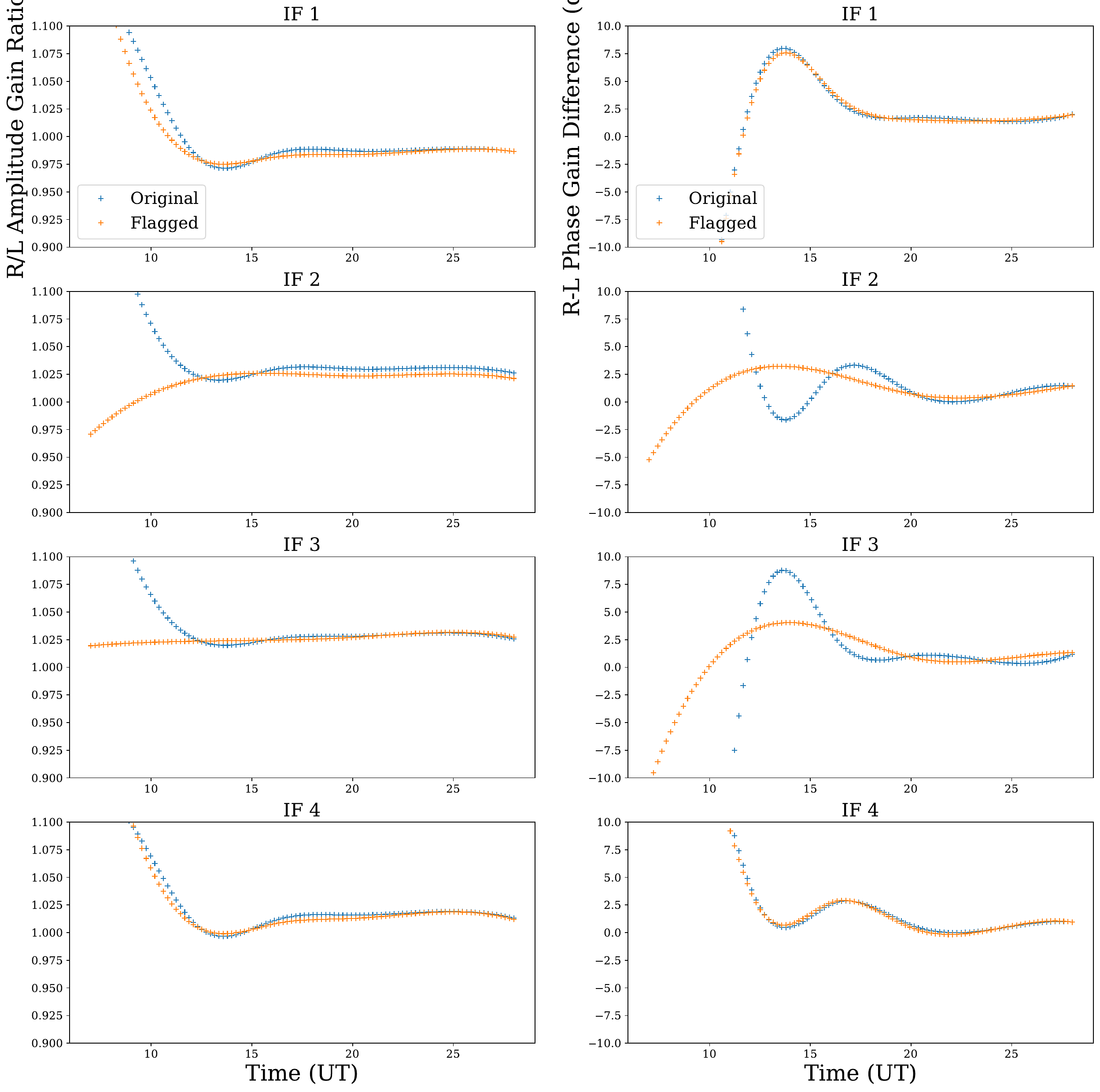}
\caption{Kitt Peak}
\end{subfigure}
\begin{subfigure}[b]{0.49\textwidth}
\includegraphics[width=\textwidth]{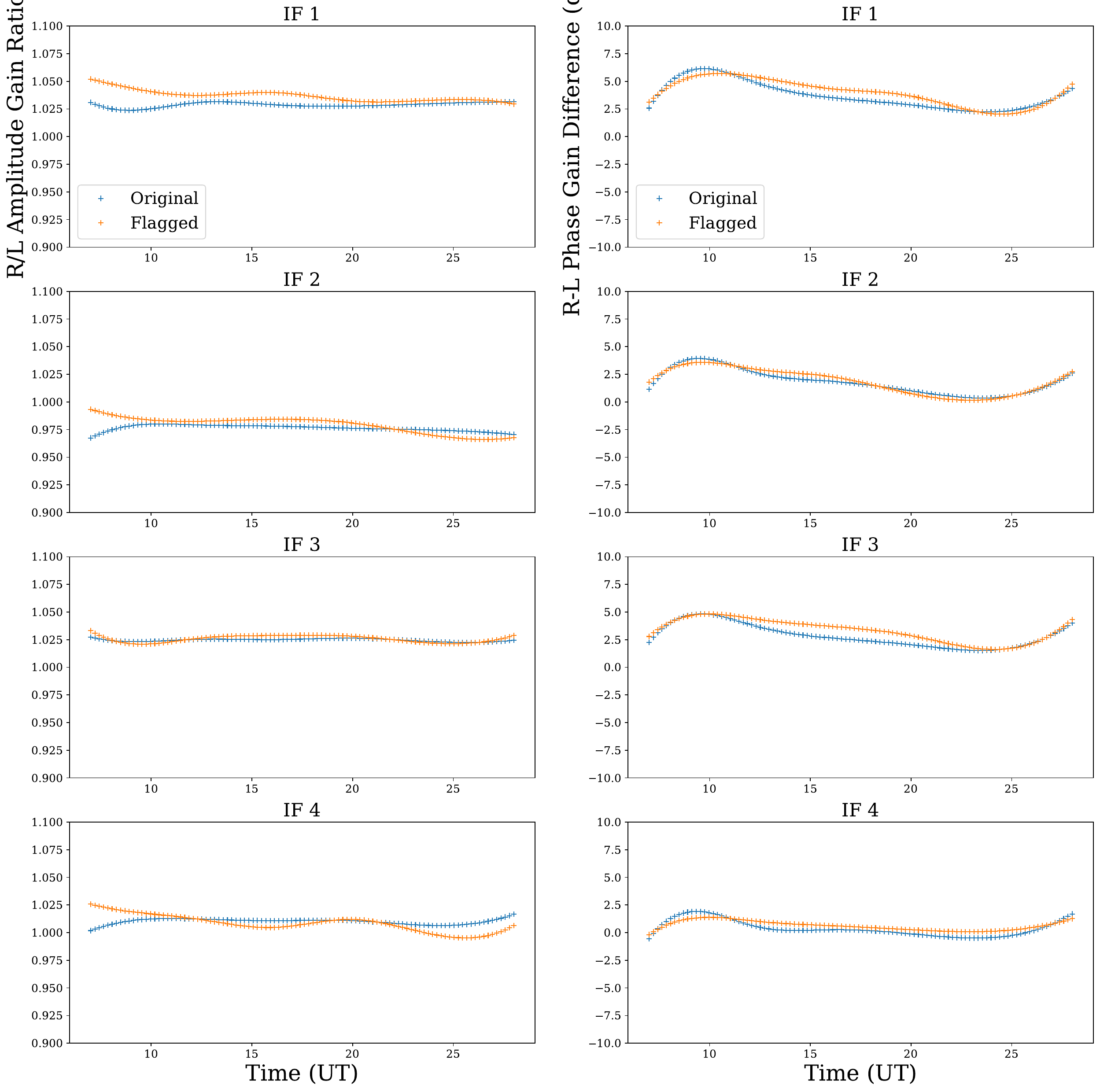}
\caption{St. Croix}
\end{subfigure}
\caption{Smoothed R/L gain curves for three stations. Shown are the gain curves obtained for the original dataset (blue) and the curves obtained when the gain transfer techniques is only applied to datasets with the long baselines flagged.}
\label{fig:rlcurve}
\end{figure*}

There are two different ways to implement the gain transfer technique, either by a running median, or by a combined, single data fidelity term. We show the gain curves obtained by us in Fig. \ref{fig:rlcurve}. In Fig. \ref{fig:CP15}, we show the circular polarization results obtained with latter technique. Here, we report in Fig. \ref{fig:CP15homan} the comparative results with the first technique (applying a running median). Both results match relatively well, with the exception of the bimodal core structures in 1243+072 and 1127+145 in which the relative importance of the corresponding positive and negative circular polarized component varies.

\begin{figure}
    \centering
    \includegraphics[width=\linewidth]{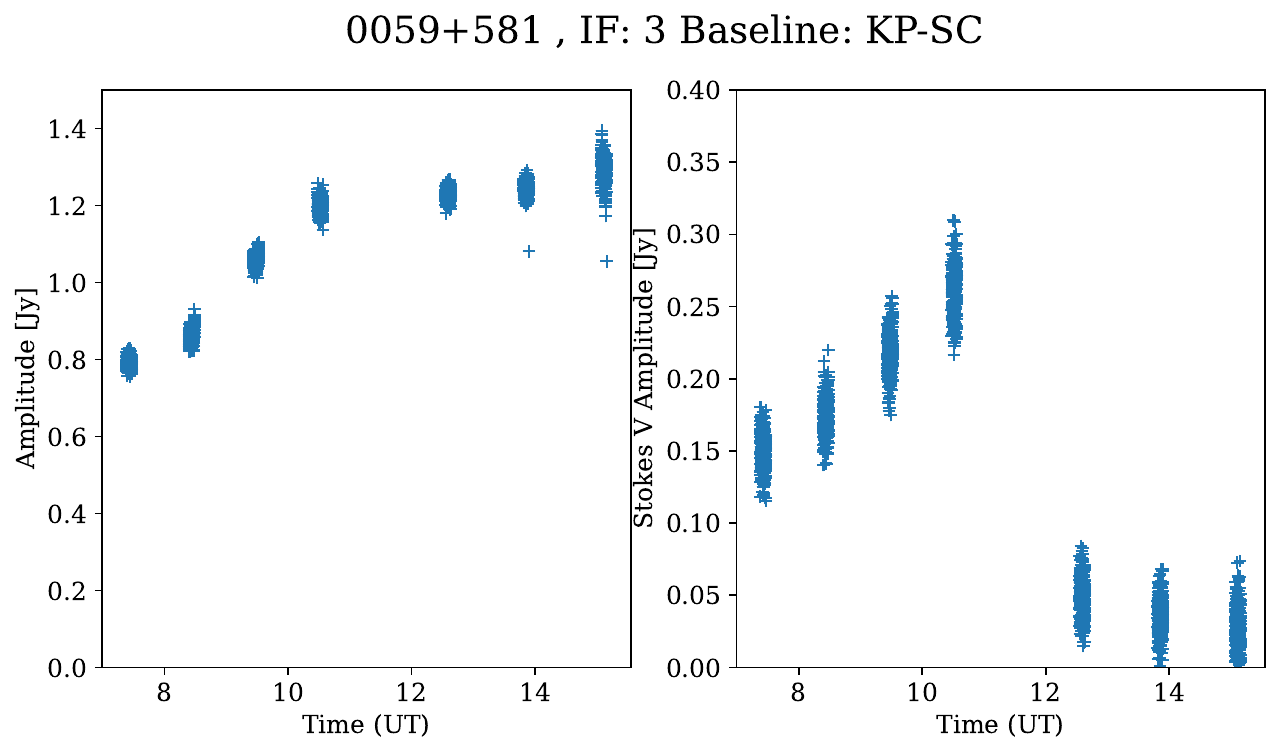}
    \caption{Amplitudes and Stokes $V$ amplitudes for IF 3 in 0059+581 on the KP-SC baseline. This is an example of a baseline that has been flagged for times earlier than UTC 12 since the gain solution was judged to be unreliable.}
    \label{fig: amplitude_jump}
\end{figure}

It has been noted that the gain curves for some antennas appear divergent towards the edges of the observing window. This is for example apparent for KP in Fig. \ref{fig:rlcurve}. This could yield potentially wrong circular polarization results and may hence be flagged during the calibration procedure. This procedure has been performed by manual inspection based on loosely defined heuristics. To this end, we examined every baseline in the self-calibrated data sets (but before application of the gain transfer) and searched for obvious jumps in the Stokes $V$ amplitude that are not visible in the amplitudes, see Fig. \ref{fig: amplitude_jump} for such an example. Such a jump alone would not necessarily indicate bad data points. The Stokes $V$ visibilities are dominated by the instrumental R/L offsets which are expected to vary over time. They are rather indicative of strong delays that need to be corrected for by the gain transfer technique. If however additionally, on one hand the baseline for which the jump occurs contains a station which gain curve shows strong variability towards the edge of the time window (e.g., KP), and on the other hand, the jump occurs at exactly the time this strong variability occurs in the gain curves (as for times before UTC 12 for KP), then we conclude that there may be delays that we cannot properly calibrate with the gain transfer technique. Data are flagged then based on a manual inspection whenever these three criteria are satisfied.

\begin{figure*}
    \centering
    \includegraphics[trim={0cm 0cm 0cm 4cm},clip,width=\textwidth]{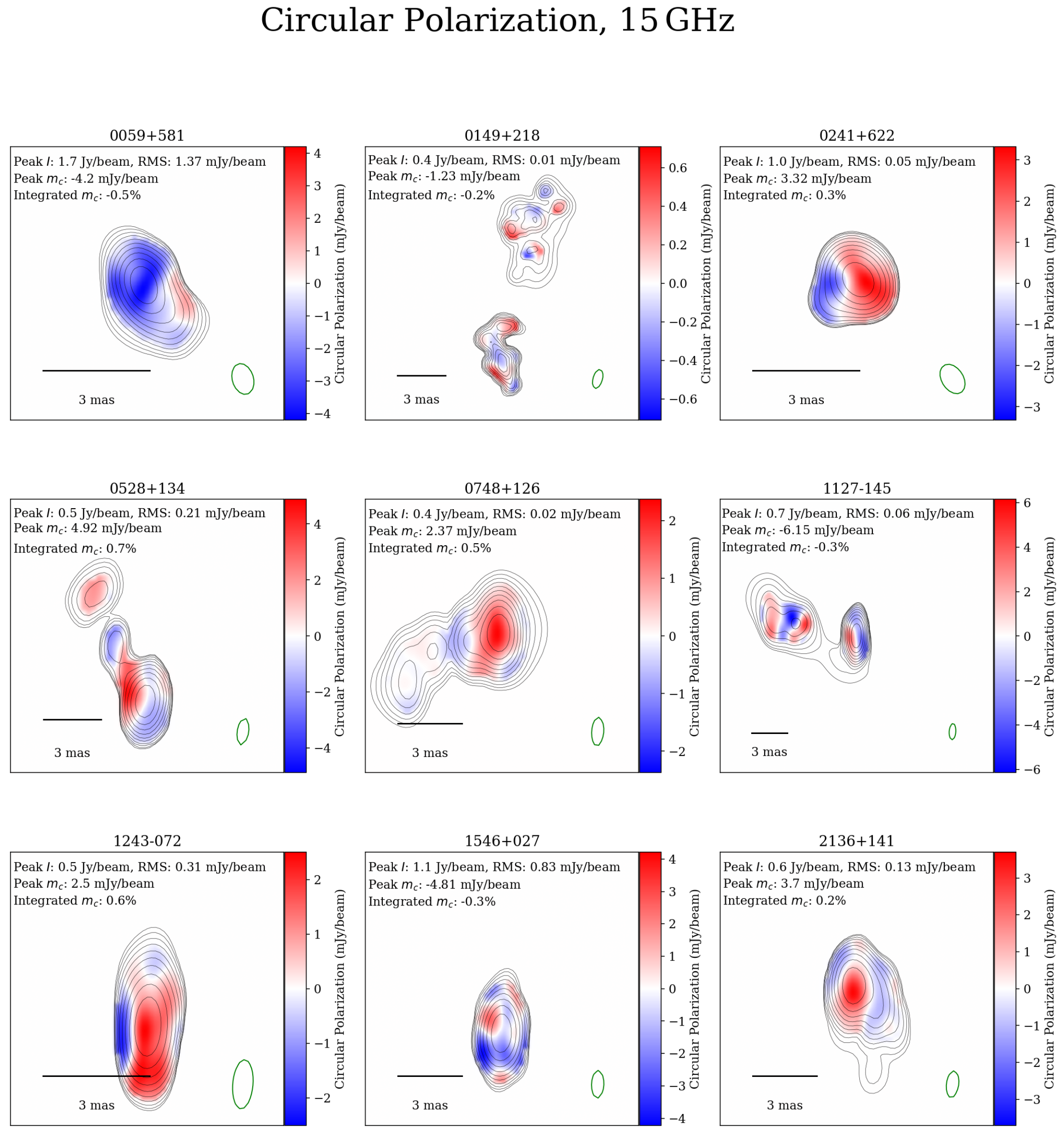}
    \caption{Calibration and Imaging with the gain-transfer technique, but the long baselines were flagged during the gain transfer.}
    \label{fig:tapered_cal}
\end{figure*}

We found that longer baselines increased circular polarization (CP) deviations from zero, influenced by self-calibration's sensitivity to small-scale structures. Obtained gain curves when flagging long baselines are shown in Fig. \ref{fig:rlcurve} as well, the final CP maps at 15\,GHz in Fig. \ref{fig:tapered_cal}. Flagging long baselines reduced this bias, producing smoother gain curves and a more balanced CP distribution, while most sources retained consistent CP structures except for 1243 and 1546, which showed significant variability due to sparse uv-coverage.

\begin{figure*}
    \centering
    \includegraphics[trim={0cm 0cm 0cm 4cm},clip,width=\textwidth]{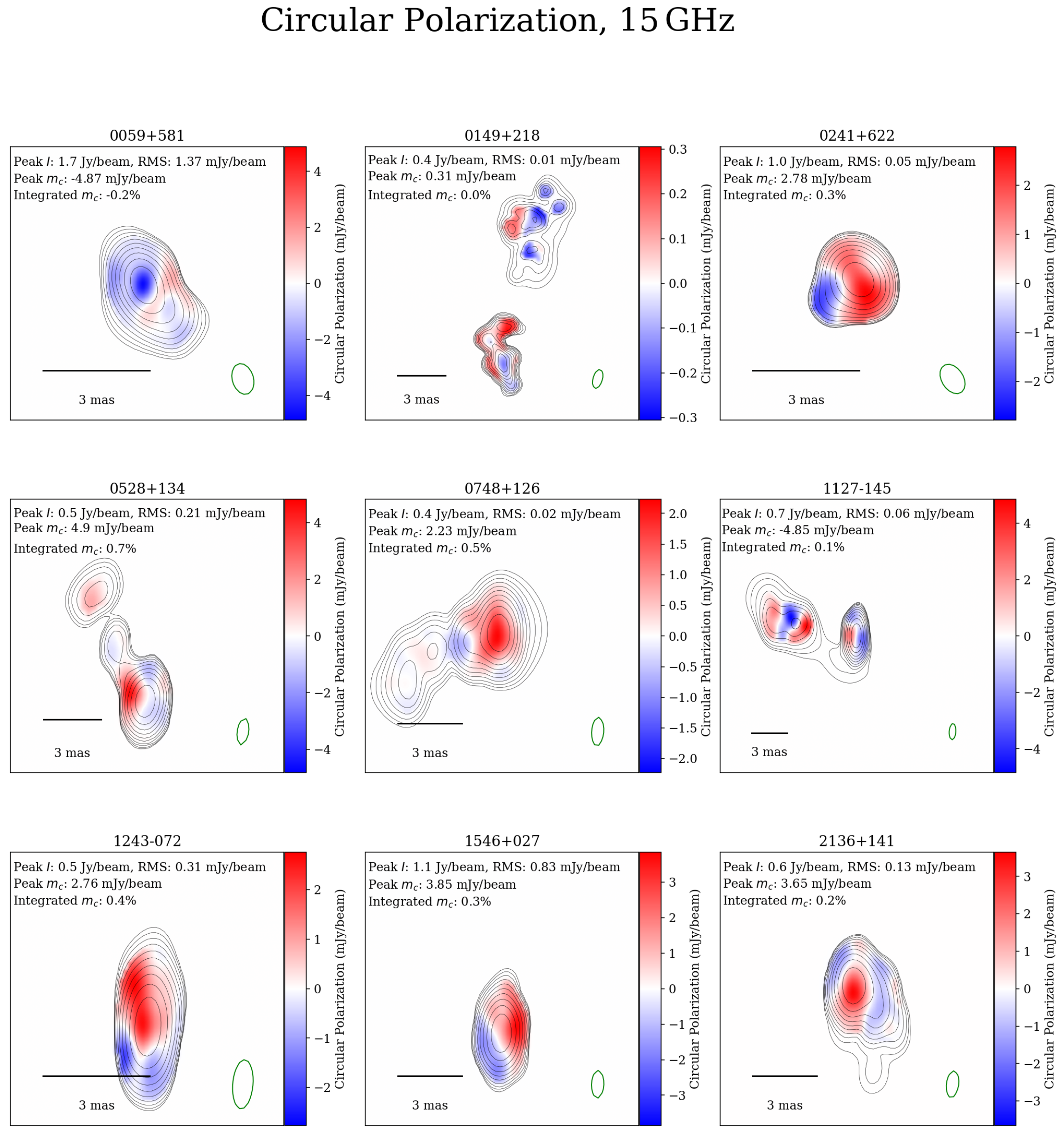}
    \caption{Results after a final self-calibration round at 15 GHz. Illustrating the same parameters as in Fig.~\ref{fig:CP15}.}
    \label{fig:CP15selfcal}
\end{figure*}

\begin{figure*}
    \centering
    \includegraphics[trim={0cm 0cm 0cm 4cm},clip,width=\textwidth]{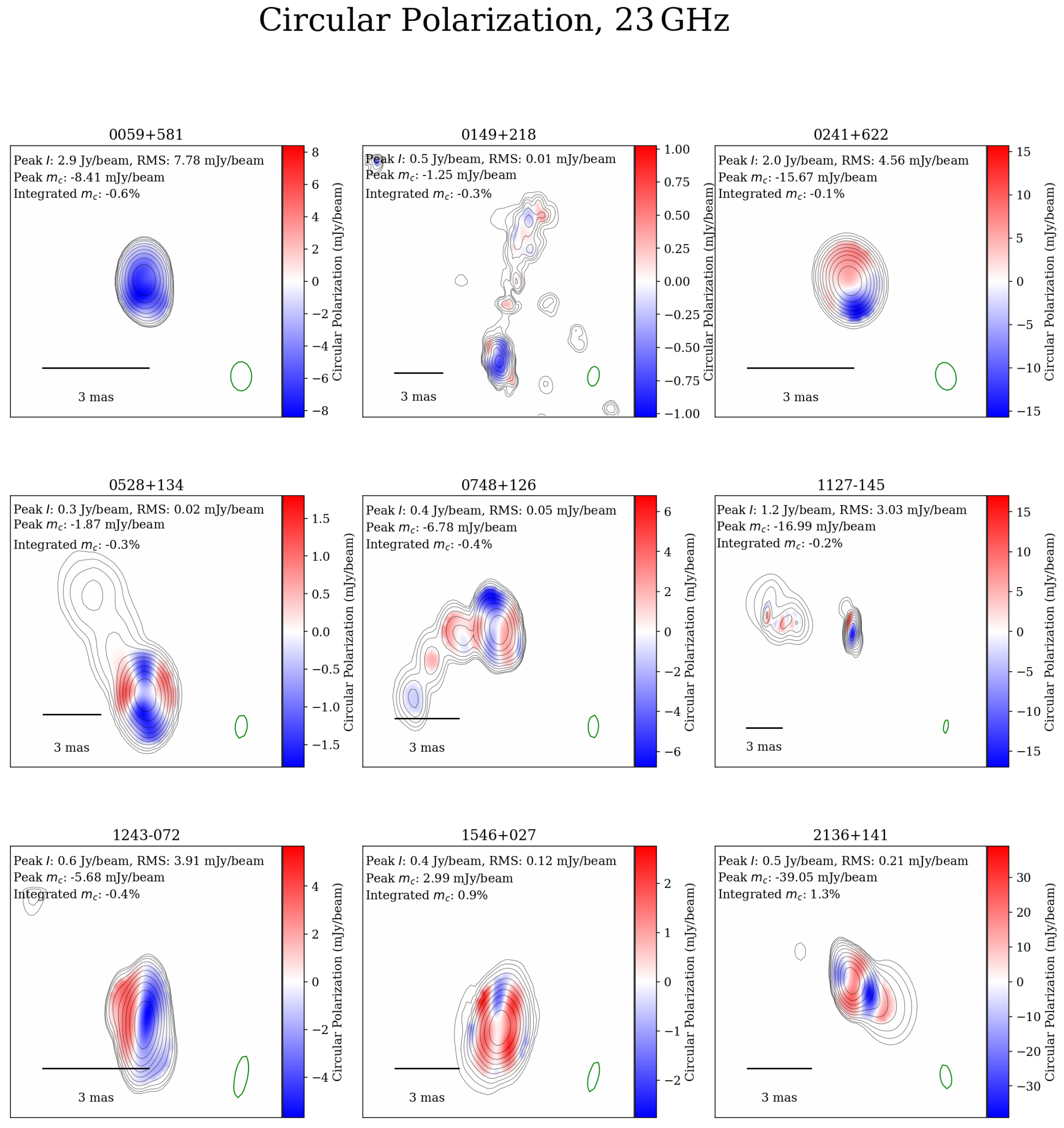}
    \caption{Results after a final self-calibration round at 23 GHz. Illustrating the same parameters as in Fig.~\ref{fig:CP23}.}
    \label{fig:CP23selfcal}
\end{figure*}

Finally, it has been suggested that antisymmetric CP could be introduced during the fringe-fitting by phase errors. This has been addressed in \citet{homan2006mojave} by a final self-calibration step in the phases. We show the reconstructions obtained by redoing this step in Fig. \ref{fig:CP15selfcal} and Fig. \ref{fig:CP23selfcal}.

\begin{figure*}
    \centering
    \includegraphics[width=\textwidth]{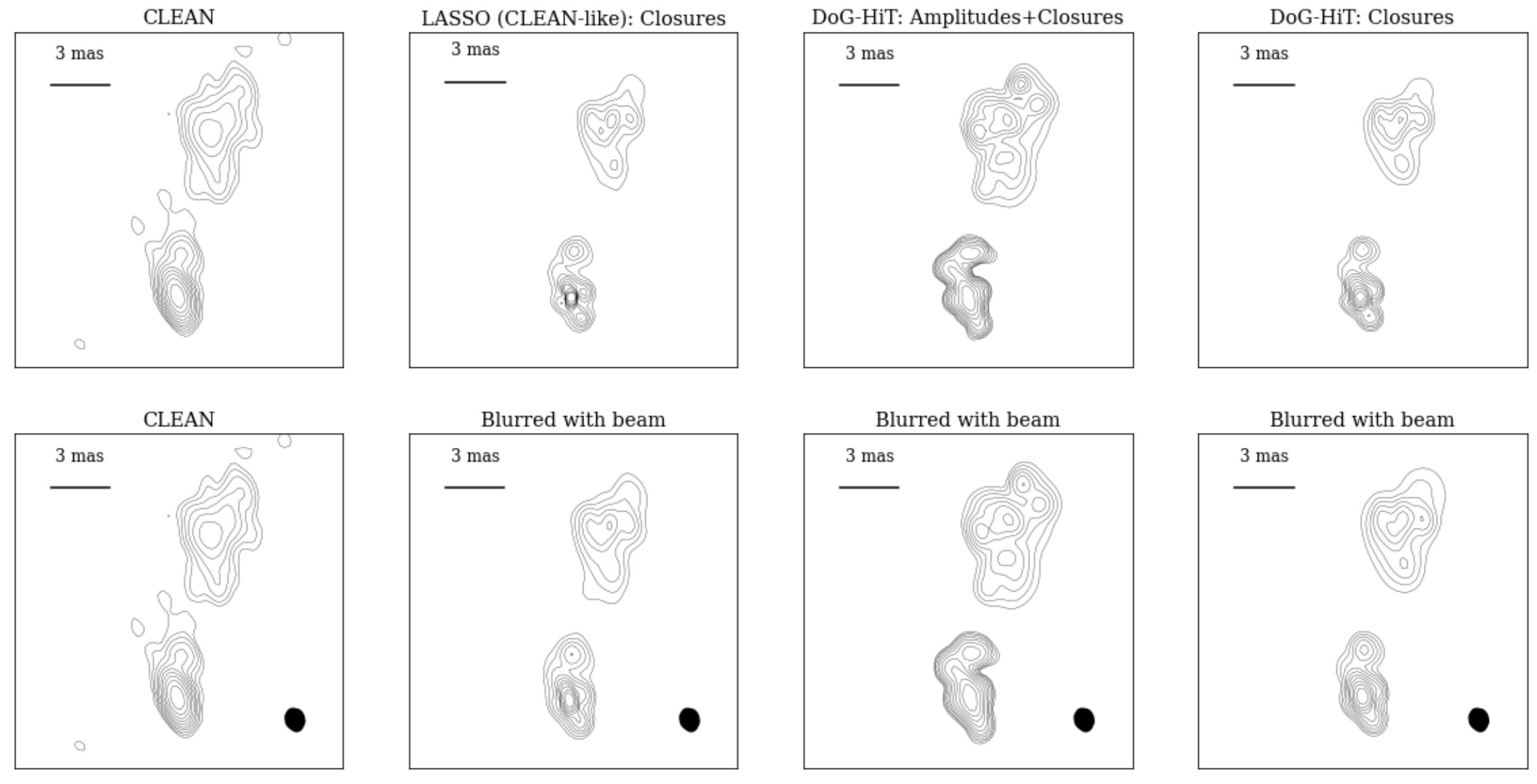}
    \caption{Reconstructions of 0149+218 with different assumptions, e.g., by CLEAN (left panel), with a LASSO scheme (CLEAN-like non-linear optimization) only fitting the closure quantities (second column), DoG-HiT fitting to amplitudes and closure quantities (third column) and DoG-HiT only fitting the closures (fourth row).}
    \label{fig:0149_I}
\end{figure*}

\begin{figure*}
    \centering
    \includegraphics[width=\textwidth]{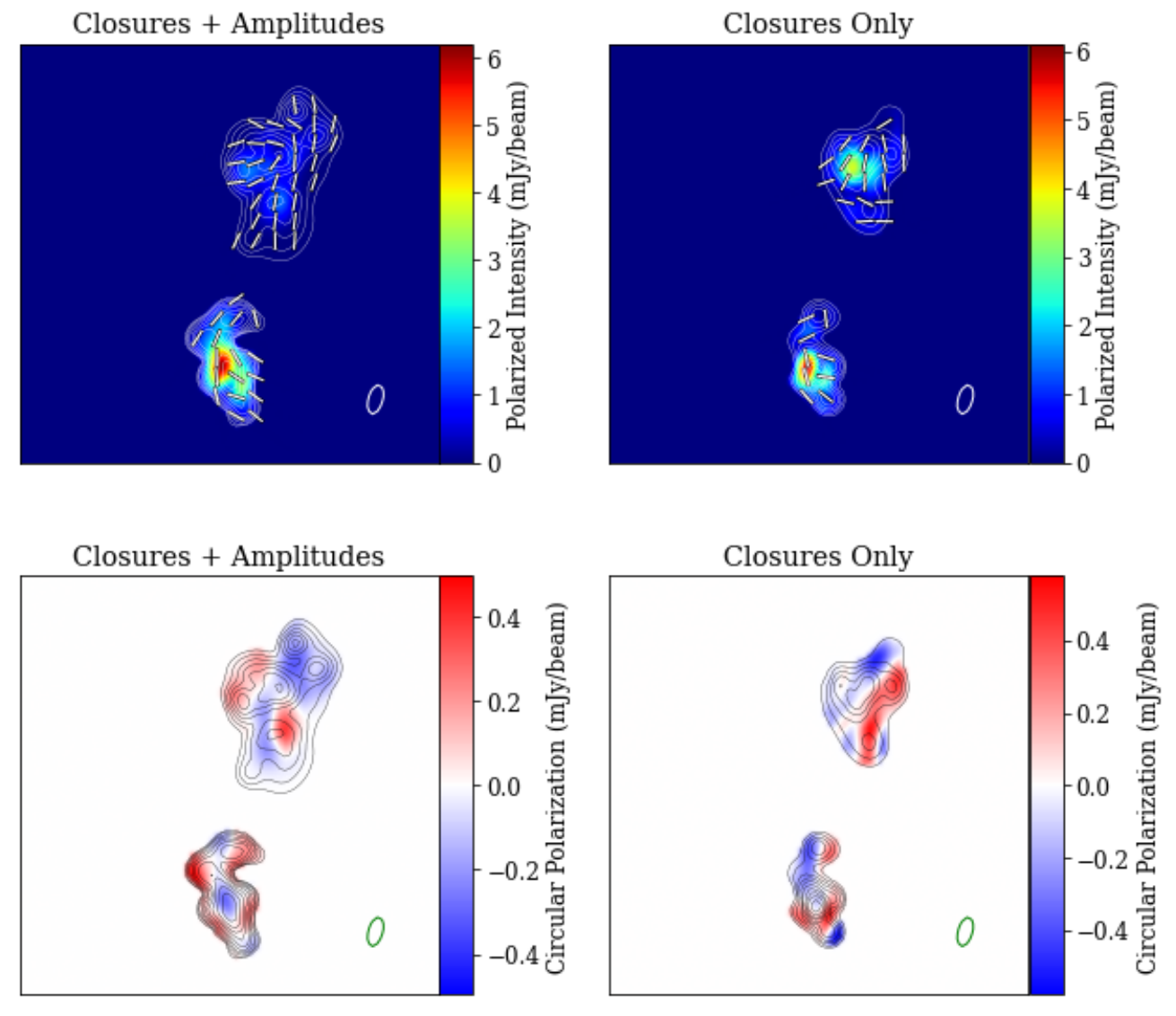}
    \caption{Polarized Reconstructions of 0149+218 either using only the closure quantities for the reconstruction in total intensity (right column), or amplitudes and closures (left column).}
    \label{fig:0149_pol}
\end{figure*}

The observed structures in the core and jet in 0149+218 do not resemble structures typically observed for this source, and with CLEAN. In Fig. \ref{fig:0149_I} and Fig. \ref{fig:0149_pol}, we present some insights into these features by redoing the imaging under varying assumptions.

\section{Impact of calibration methods on 1243$-$072 and 1546+027}\label{app:ex}
Our approach for the calibration circular polarization has also been validated in App.~\ref{app:comparison} using an alternative gain estimation technique (running median rather than a smooth curve) and with an additional self-calibration step. For most sources, the relative structures remain consistent. Notably, in resolved double structures, the relative brightness varies. We observe two exceptions to a consistent result, namely, 1243$-$072 and 1546+027. That is, across all the tests conducted, the two sources, 1243 and 1546, consistently exhibited the most significant variations between methods. This strongly suggests that the circular polarization signal in these two sources may be influenced by the gain calibration process.
In line with this, we have marked the estimates for these two sources in Tabs.~\ref{tab:source_list} \& \ref{tab:pol} and have effectively excluded them from the broader interpretation.
Upon inspecting the $uv$-coverage, this observation makes sense: 1243$-$072 and 1546+027 have the sparsest coverage and are the most weakly constrained sources in the dataset. Additionally, we would like to highlight an interesting aspect of this comparison. The recovered structure in 1243$-$072, when flagging long baselines, appears more similar to the reconstructions obtained using a running median than to those derived from our gain calibration approach. This may further indicate that our gain calibration method is more sensitive to the longer baselines than traditional techniques.

\end{appendix}
\end{document}